%% file: main.tex
\documentclass[%
 reprint,
superscriptaddress,
nofootinbib,
 amsmath,amssymb,
 aps,prd,
floatfix,
]{revtex4-2}

\usepackage{graphicx}
\usepackage{dcolumn}
\usepackage{bm}
\usepackage{hyperref}
\usepackage[mathlines]{lineno}
\usepackage{booktabs,array}
\usepackage{longtable}
\usepackage{multirow}
\usepackage{upgreek}

\usepackage{xcolor}

\newcommand{\h}{$h^{+}$}
\newcommand{\e}{$e^{-}$}
\newcommand{\eh}{$e^{-}h^{+}$}
\newcommand{\neh}{$n_{\textrm{eh}}$}

\begin{document}


\title{Improved Modelling of Detector Response Effects in Phonon-based Crystal Detectors used for Dark Matter Searches}

\input{authors}

\date{\today}

\begin{abstract}
Various dark matter search experiments employ phonon-based crystal detectors operated at cryogenic temperatures. Some of these detectors, including certain silicon detectors used by the SuperCDMS Collaboration, are able to achieve single-charge sensitivity when a voltage bias is applied across the detector. The total amount of phonon energy measured by such a detector is proportional to the number of electron-hole pairs created by the interaction. However, crystal impurities and surface effects can cause propagating charges to either become trapped inside the crystal or create additional unpaired charges, producing non-quantized measured energy as a result. A new analytical model for describing these detector response effects in phonon-based crystal detectors is presented. This model improves upon previous versions by demonstrating how the detector response, and thus the measured energy spectrum, is expected to differ depending on the source of events. We use this model to extract detector response parameters for SuperCDMS HVeV detectors, and illustrate how this robust modelling can help statistically discriminate between sources of events in order to improve the sensitivity of dark matter search experiments.

\end{abstract}

\maketitle

\section{\label{sec:intro}Introduction}
Cryogenic solid-state detectors are used in a number of dark matter (DM) search experiments \cite{0vev, cpd, cresst_10ev, cdms_lite, cresst, edelweiss}. In these experiments, incoming DM particles are expected to scatter off of the detector nuclei or electrons, creating phonon signals which are measured by high resolution phonon sensors. Resolution on the order of 1\,eV is achieved, which allows for reduced energy thresholds and enables the detection of nuclear recoils with energies as low as $\sim$10\,eV \cite{0vev, cpd, cresst_10ev}. Low-mass DM candidates that produce small interaction energies can be probed via electron recoils by measuring the ionization signal --- the number of produced \eh{} pairs in the detector \cite{HVeVR1, HVeVR2}. In phonon-based crystal detectors, when a voltage bias is applied across the crystal, the ionization signal is converted into an amplified phonon signal via the Neganov-Trofimov-Luke (NTL) effect \cite{ntl1, ntl2}. A charge carrier with a charge $e$ accelerated by the electric field scatters off of the crystal lattice and produces NTL phonons with the total energy equal to the work done by the electric field to move the charge through the electric potential difference $\Delta\varphi$:
\begin{equation}
    E_\mathrm{NTL} = e\,\Delta\varphi.
\end{equation}
Normally, when an \eh{} pair is created in the crystal, each charge drifts in the electric field all the way to the corresponding electrode on the crystal surface. Together they traverse the entire voltage bias of the detector, so the total energy of the produced NTL phonons is given by:
\begin{equation}
    E_\mathrm{NTL} = n_{\textrm{eh}} e V_\mathrm{bias},
\end{equation}
where $n_\mathrm{eh}$ is the number of \eh{} pairs and $V_\mathrm{bias}$ is the voltage bias. The total phonon energy produced in an event is then given by:
\begin{equation}\label{eq:phonon_energy}
    E_{\textrm{ph}} = E_{\textrm{dep}} + n_{\textrm{eh}} e V_{\textrm{bias}},
\end{equation}
where $E_{\textrm{dep}}$ is the energy deposited in the detector by the incoming particle. For a detector with a good phonon energy resolution $\sigma_{\textrm{res}}$ and a large voltage bias, where $\sigma_{\textrm{res}} \ll eV_{\textrm{bias}}$, the spectrum of the phonon energy in Eq.~\ref{eq:phonon_energy} is expected to have quantized peaks corresponding to the integer number of created \eh{} pairs. This \eh{}-pair quantization is observed in SuperCDMS high-voltage (HV) and HV eV-scale (HVeV) detectors when operated with a voltage bias on the order of 100\,V~\cite{HVeVR1,HVeVR2,Ren:2021}.

Due to the presence of impurities in the crystal, a non-quantized amount of NTL energy can be produced in an event. We distinguish two categories of effects causing non-quantized NTL energy: charge trapping (CT) and impact ionization (II). In a CT process, a charge carrier gets trapped in an impurity state in the bulk of the crystal. In an II process, a propagating charge ejects  (or ``ionizes") an additional unpaired charge from a shallow impurity state. Trapped charge carriers and unpaired charge carriers created in an II process terminate or start their trajectories in the bulk of the detector, respectively. As a result, they traverse only a fraction of the voltage bias, producing a non-quantized amount of NTL energy.

A proper modelling of these detector response effects is crucial for DM search analyses. In Sec.~\ref{sec:exp_model}, we develop an analytical model (the so-called ``exponential CTII" model) that describes the NTL energy spectrum for events affected by the CT and II processes. We improve upon the previously used CT and II model introduced in Ref.~\cite{Ponce2020} (the so-called ``flat CTII" model) by taking into account the distribution of locations at which the CT and II processes occur. We demonstrate a difference between the NTL energy spectra of events produced on the detector surface and events produced in the detector bulk that can be used for statistical discrimination between surface background and bulk DM events. In Sec.~\ref{sec:det_response}, we incorporate the CT and II model into the full detector response model, and take into account additional surface effects that may be relevant to certain calibration data. This modelling is used in Sec.~\ref{sec:results} to extract detector response parameters for HVeV detectors.

\section{\label{sec:exp_model}Exponential CTII Model}
The underlying physical assumption of the exponential CTII model is that there are three possible processes that can occur to a charge carrier (an electron or a hole) when it traverses the bulk of the crystal under the influence of an electric field. It can get trapped in an impurity state, it can create a single free electron from an impurity state by promoting it into the conduction band, or it can promote an electron from the valence band to an impurity state, creating a single hole in the valance band. The probabilities for these processes to occur may differ between holes and electrons; therefore we consider in the model a total of six different CT and II processes: electron trapping (``CTe''), hole trapping (``CTh''), creation of a hole by an electron (``IIeh''), creation of an electron by an electron (``IIee''), creation of an electron by a hole (``IIhe'') and creation of a hole by a hole (``IIhh'').

The model assumes that each of the six processes has a small constant probability of occurring at any point of the charge carrier's trajectory, independent of the location in the bulk of the crystal, of the path already travelled by the charge, and the presence of other charges simultaneously traversing the crystal. Additionally, it is assumed that charges propagate along some $z$ axis that is parallel to a uniform electric field (detectors, including the HVeV detectors used in Refs.~\cite{HVeVR1,HVeVR2}, are typically designed to have a uniform electric field throughout the bulk). We start by considering that impurities are distributed uniformly throughout the bulk of the crystal, where we let $p_{i}$ denote the probability for a charge to undergo a certain process $i$ per unit of distance travelled along the $z$ axis. Here, $i$ refers to the specific CT or II process a charge may undergo (CTe, CTh, IIee, IIeh, IIhe, or IIhh). $p_{i}$ itself may depend on various factors, including the impurity density and the amount of charge diffusion.  If a charge travels a distance $\Delta z$ in $n$ steps, the total probability of the charge \textit{not} undergoing some CT or II process $\overline{C}_{i}(\Delta z)$ is $(1 - p_{i}\Delta z/n)^n$. In the limit of infinitesimally small step sizes, $\overline{C}_{i}(\Delta z)$ becomes:

\begin{equation}\label{eq:Cbar_z}
\begin{split}
     \overline{C}_{i}(\Delta z) &= \lim_{n \to \infty} \left (1 -  p_{i}\frac{\Delta z}{n} \right)^{n} \\
     &= e^{-p_{i} \Delta z} \\
     &= e^{-\Delta z/\tau_{i}},
\end{split}
\end{equation}
where the $p_{i}$ term in Eq.~\ref{eq:Cbar_z} is replaced with $1/\tau_{i}$, with $\tau_{i}$ defining the characteristic length of that particular CT or II process. $\overline{C}_{i}(\Delta z)$ is the complementary cumulative distribution function of the probability density function (PDF) that describes the probability for a charge to travel a distance $\Delta z$ before a particular CT or II process occurs. This PDF is therefore given by:

\begin{equation}\label{eq:Pi_z}
\begin{split}
    P_{i}(\Delta z) &= \frac{\mathrm{d}}{\mathrm{d}(\Delta z)} (1 - \overline{C}_{i}(\Delta z)) \\
    &= \frac{\mathrm{d}}{\mathrm{d}(\Delta z)} \left ( 1 - e^{-\Delta z/\tau_{i}} \right) \\
    &= \frac{1}{\tau_{i}}e^{-\Delta z/\tau_{i}}.
\end{split}
\end{equation}

While these PDFs are described in terms of a distance travelled, the model also imposes the condition that the charges terminate when reaching the crystal surface. That means for a crystal with a thickness $Z$, the charges are bound between $z=0$ and $z=Z$. For convenience we choose $Z=1$, and let $z$ describe the proportion of the crystal thickness rather than a physical distance. The six characteristic lengths $\tau_i$ measured in fractions of the crystal thickness are the only fundamental input parameters of the model. We write these characteristic lengths in terms of probabilities $f_{i}$ defined as:
\begin{equation}\label{eq:p_tau}
    f_i \equiv \int_0^1 P_i(\Delta z)\mathop{\mathrm{d}(\Delta z)} = 1 - e^{-1/\tau_i}.
\end{equation}

Hence $f_{i}$ is the probability of a particular process occurring if a charge can traverse the entire length of the detector. Equations~\ref{eq:Pi_z} and~\ref{eq:p_tau} are repeated for each of the six processes, and together make up the fundamental building blocks of our exponential CTII model. 

The end product of the model is a PDF of the NTL energy produced in an event. This energy is proportional to the distance travelled by the charges along the field lines. We adopt an energy scale $E_{\textrm{neh}}$ such that a unit of $E_{\textrm{neh}}$ is equivalent to the amount of NTL energy produced by a charge that travels a distance equal to the thickness of the crystal. Using this energy scale, a charge going from $z=0$ to $z=1$, as well as an \eh{} pair starting at $z=0.5$ whereby both charges travel a distance $\Delta z = 0.5$, will result in a total energy of  $E_{\textrm{neh}} = 1$. With such units, there is a one-to-one correspondence between the PDFs of the total NTL energy and the total distance travelled by the charges along the electric field.

The exponential CTII model is constructed by finding the analytical solutions for the NTL energy produced by a single \eh{} pair for events of three distinct classes. The first are surface events, where a single charge is created at one of the surfaces (i.e. along the $z=0$ or $z=1$ plane) and propagates toward the opposite surface; this class of events does not include events created along the lateral surfaces of the crystal. Surface events correspond to laser or light-emitting diode (LED) calibration data, whereby optical photons are absorbed near the $z=0$ or $z=1$ surface of the crystal, as well as to charge leakage originating at the crystal surface. The second class of events are single charges produced throughout the bulk of the crystal. These events may correspond to some charge leakage process that happens throughout the detector bulk. The third class of events are bulk-\eh{} pairs produced throughout the bulk of the crystal. These events are what is expected for DM interactions. For each class of events, we consider various unique combinations of CT and II processes occurring to the charges, and solve for the probabilities of measuring an energy of $E_{\textrm{neh}}$ given those unique combinations of processes.

Modelling multiple II processes in a single event poses a significant challenge: each additional II process allowed adds an new charge carrier, causing the number of potential combinations of CT and II processes to grow exponentially, and the complexity of each new solution greatly increases. For this reason, we limit the number of solutions to a certain ``order" of processes, where the order of a process is defined as follows: for processes of order $N$, charges that participated or were produced in a primary II process can take part in no more than $(N-1)$ additional II processes. For surface events and bulk-single-charge events, the solutions for processes up to second order are found, resulting in 28 unique solutions for each event type. For bulk-\eh{}-pair events, the solutions for processes up to first order are found, resulting in 16 unique solutions. When solving for these analytical solutions, we assume that any charges existing after the order limit is reached will propagate to a crystal surface with 100\,\% probability. Appendix~\ref{app:single_eh_solution} provides a detailed description of how these solutions are found, with Appendix~\ref{sssec:surface_Events}, Appendix~\ref{sssec:bulk_singlecharge_Events}, and Appendix~\ref{sssec:bulk_neh_Events} adding further details on solving the solutions for each of the three classes of events.  The full list of process combinations and the corresponding solutions are catalogued in the Supplemental Material and are displayed in Fig.~\ref{fig:1eh_solutions}. It is immediately apparent how the computed PDFs differ for the different classes of events. Namely, the regions above and below the first \eh{}-pair peak are relatively flat for surface events, in contrast to bulk-\eh{}-pair events where the PDF in the same regions is more curved. Furthermore, the PDF for bulk-single-charge events does not have a delta function at $E_{\mathrm{neh}} = 1$. 

\begin{figure}[t!]
\begin{center}
\includegraphics[width=1.0\columnwidth]{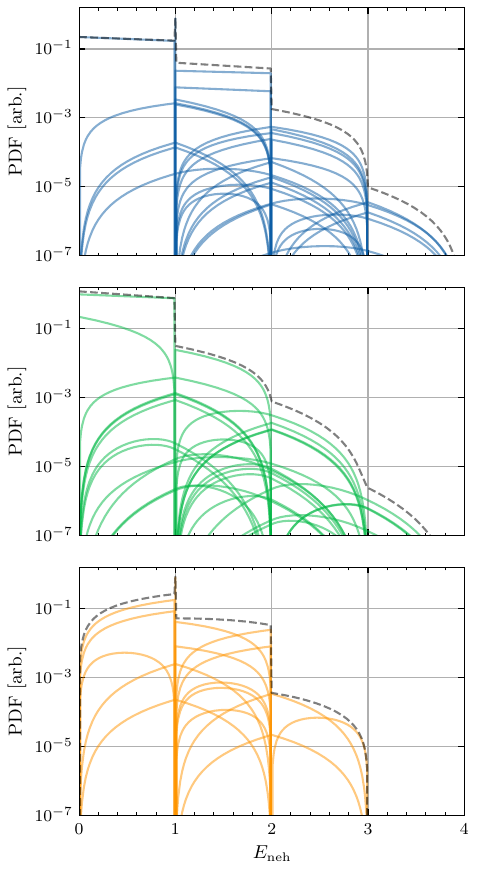}
\end{center}
\caption[Single electron-hole-pair solutions]{Analytical solutions in the $E_{\textrm{neh}}$ energy space of the exponential CTII model for single-\eh{}-pair events. The unique solutions represented by the solid, coloured curves are found for surface events (top), bulk-single-charge events (middle), and bulk-\eh{}-pair events (bottom). This example is shown for arbitrary values of the CT and II parameters: $f_{\mathrm{CTe}} = 20$\,\%, $f_{\mathrm{CTh}} = 10$\,\%, $f_{\mathrm{IIee}} = 1$\,\%, $f_{\mathrm{IIeh}} = 3$\,\%, $f_{\mathrm{IIhe}} = 1$\,\%, and $f_{\mathrm{IIhh}} = 5$\,\%, and the top and middle plots assume that the initial charge is an electron. The black, dashed curves in each plot are the sums of the analytical solutions for each event type and are examples of $F^{(1)}_{\textrm{type}}(E_{\textrm{neh}})$, the one-\eh{}-pair PDF.}
\label{fig:1eh_solutions}
\end{figure}

The analytical solutions are found for when there is initially only a single charge or \eh{} pair produced. However large energy depositions in the crystal will often produce multiple charges or \eh{} pairs for a single event. Let $F^{(1)}_{\textrm{type}}(E_{\textrm{neh}})$ be the probability distribution function for one charge or \eh{} pair in the \neh{}-energy space. The ``type" refers to the specific event type to model: either surface events, bulk-single-charge events, or bulk-\eh{}-pair events. $F^{(1)}_{\textrm{type}}(E_{\textrm{neh}})$ is found by summing the analytical solutions for the given event type, and examples of this function are shown by the black, dashed curves in Fig.~\ref{fig:1eh_solutions}. Without any additional detector response, the PDF for $j$ \eh{} pairs $F^{(j)}_{\textrm{type}}(E_{\textrm{neh}})$ is found by convolving $F^{(1)}_{\textrm{type}}(E_{\textrm{neh}})$ with itself $(j-1)$ times:

\begin{equation}\label{eq:neh_ind_pdf}
    F^{(j)}_{\textrm{type}}(E_{\textrm{neh}}) = F^{(j-1)}_{\textrm{type}}(E_{\textrm{neh}}) \ast F^{(1)}_{\textrm{type}}(E_{\textrm{neh}}).
\end{equation}

\begin{figure}[t!]
\begin{center}
\includegraphics[width=1.0\columnwidth]{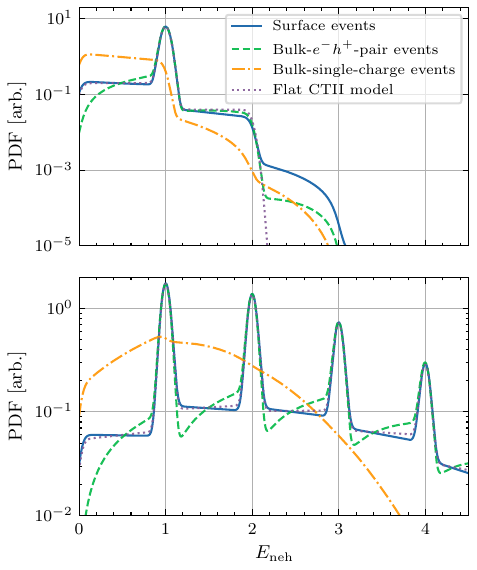}
\end{center}
\caption[Single and multi electron-hole-pair solutions]{Example PDFs found for single-\eh{}-pair events $F^{(1)}_{\textrm{type}}(E_{\textrm{neh}})$ (top) and multi-\eh{}-pair events $H(E_{\textrm{neh}})$ (bottom). The PDFs are computed for surface events (solid, blue curves), bulk-\eh{}-pair events (dashed, green curves), and bulk-single-charge events (dot-dash, orange curves) using the exponential CTII model. For comparison, the PDFs computed using the flat CTII model from Ref.~\cite{Ponce2020} are shown by the dotted, purple curves. These examples are shown for arbitrary CT and II parameters: $f_{\mathrm{CTe}} = f_{\mathrm{CTh}} = 20$\,\% and $f_{\mathrm{IIee}} = f_{\mathrm{IIeh}} = f_{\mathrm{IIhe}} =f_{\mathrm{IIhh}} = 2$\,\%; for the flat CTII model, $f_{\mathrm{CT}} = 20$\,\% and $f_{\mathrm{II}} = 4$\,\%. Furthermore, the multi-\eh{}-pair solutions assume that the $a_{j}$ terms in Eq.~\ref{eq:multi_eh_pdf} that describe the \eh{}-pair probabilities follow a Poisson distribution with a mean of two \eh{} pairs. For illustrative purposes, the PDFs are convolved with a Gaussian function with a width of $E_{\textrm{neh}} = 0.05$ to emulate the detector energy resolution.}
\label{fig:single-multi-CTII}
\end{figure}

In practice, $F^{(j)}_{\textrm{type}}(E_{\textrm{neh}})$ is found using numerical convolution. We can use this to construct the PDF for events that generate multiple \eh{} pairs, defined as $H(E_{\textrm{neh}})$. The solution for $H(E_{\textrm{neh}})$ up to $J$ \eh{} pairs is given by:

\begin{equation}\label{eq:multi_eh_pdf}
    H(E_{\textrm{neh}}) =  \sum_{j=1}^{J} a_{j} \cdot F^{(j)}_{\textrm{type}}(E_{\textrm{neh}}),
\end{equation}
where $a_{j}$ are the weights associated with producing $j$ \eh{} pairs, which are discussed more in Sec.~\ref{sec:det_response}. A comparison of the PDFs for single-\eh{}-pair events $F^{(1)}_\textrm{type}(E_{\textrm{neh}})$ and multi-\eh{}-pair events $H(E_{\textrm{neh}})$ for different event types is shown in Fig.~\ref{fig:single-multi-CTII} for arbitrary CT and II probabilities. The PDFs are convolved with a Gaussian function to emulate the energy resolution for illustrative purposes. Furthermore, the solutions are compared to the PDFs computed using the flat CTII model described in Ref.~\cite{Ponce2020}.

The example PDFs from Fig.~\ref{fig:single-multi-CTII} allow us to make some broad observations about the exponential CTII model. First, while the higher-order processes are significant for the single-\eh{}-pair solutions (as seen in the top plot of Fig.~\ref{fig:single-multi-CTII} above $E_{\mathrm{neh}} = 2$), they generally become less significant or even negligible for multi-\eh{}-pair solutions. Second, the type of events being modelled has a significant impact on the shape of the PDFs between the \eh{}-pair peaks. Notably, the between-peak shape for the bulk-\eh{}-pair events differs greatly from that of surface events, as well as that of the flat CTII model which does not differentiate between event types. 

\section{\label{sec:det_response}Extended Detector Response Model}
\subsection{Single- and Multi-Hit Solutions}\label{ssec:extended_model_singlemulti}
Equation~\ref{eq:multi_eh_pdf} describes the PDF of producing a certain amount of NTL energy for a given event type that generates multiple \eh{} pairs, $H(E_{\textrm{neh}})$, which is derived from the analytical solutions of the exponential CTII model. However $H(E_{\textrm{neh}})$ can be extended to include other detector response effects that are either measured or expected. These additional effects include ionization probabilities, conversion to the phonon energy scale, and continuous spectra of energy deposition. To start, we define $H^{(1)}$ as the PDF for events resulting from a single interaction of a particle with the crystal, which we call ``single-hit" events. Examples of a single-hit event include a single photon absorbed by the crystal, or a single DM particle scattering off of an electron. First we will construct a general formula for $H^{(1)}$, and then subsequently see how this formula is used to model specific interactions from various sources.

The first step to extend the detector response model is to replace the generic weights $a_{j}$ in Eq.~\ref{eq:multi_eh_pdf} with the probability mass function (PMF) describing the probability of producing a given amount of ionization. The ionization PMF is specific to the detector material, and is a function of the energy deposited in the detector $E_{\textrm{dep}}$. Let $p_{\textrm{eh}}(j\,|\,E_{\textrm{dep}})$ describe the ionization probability of producing $j$ \eh{} pairs given $E_{\textrm{dep}}$. For silicon, results of the ionization yield at low energies can be found in Ref.~\cite{Ramanathan2020}.

Next we need to convert the PDFs to the correct energy scale. As mentioned in Sec.~\ref{sec:intro}, event energies are measured by the total phonon energy $E_{\textrm{ph}}$ described by Eq.~\ref{eq:phonon_energy}, whereas the $F^{(1)}_{\textrm{type}}$ functions of the exponential CTII model are described in the \neh{}-energy space $E_{\textrm{neh}}$. Using Eq.~\ref{eq:phonon_energy}, $E_{\textrm{neh}}$ can be written in terms of $E_{\textrm{ph}}$ as:
\begin{equation}\label{eq:Eneh_to_Eph}
E_{\textrm{neh}} = \frac{E_{\textrm{ph}} - E_{\textrm{dep}}}{eV_{\textrm{bias}}}.
\end{equation}

This change in energy units also changes the overall scaling of the PDFs. To account for this, the PDFs must be scaled by a factor of $\left |\mathrm{d}E_{\textrm{neh}}/\mathrm{d}E_{\textrm{ph}} \right | = 1/eV_{\textrm{bias}}$. Finally, we need to consider the general case where there is a continuum of energy depositions that can occur for a given source of events. This continuum can be described by a \textit{normalized} differential rate spectrum $\mathrm{d}\overline{R}/\mathrm{d}E_{\textrm{dep}}(E_{\textrm{dep}})$, where for a total single-hit event rate of $R_{\textrm{tot}}$, $\mathrm{d}\overline{R}/\mathrm{d}E_{\textrm{dep}}(E_{\textrm{dep}}) = 1/R_{\textrm{tot}} \cdot \mathrm{d}R/\mathrm{d}E_{\textrm{dep}}(E_{\textrm{dep}})$. Putting this all together, the extended detector response model for single-hit events in the phonon energy space modelled up to $J$ \eh{} pairs is given as:
\begin{align}\label{eq:H1_extended}
\begin{split}
H^{(1)}(E_{\textrm{ph}}) &= \sum_{j=1}^{J} \Biggl(\int_{0}^{\infty} \mathrm{d}E_{\textrm{dep}} \, p_{\textrm{eh}}(j\,|\,E_{\textrm{dep}}) \, \times \\
&\qquad \frac{F^{(j)}_{\textrm{type}} \left (\frac{E_{\textrm{ph}} - E_{\textrm{dep}}}{eV_{\textrm{bias}}} \right )}{eV_{\textrm{bias}}} \frac{\mathrm{d}\overline{R}}{\mathrm{d}E_{\textrm{dep}}}(E_{\textrm{dep}}) \Biggr).
\end{split}
\end{align}

Here we assume that $J$ is large enough such that the ionization PMF sums to unity for all $E_{\textrm{dep}}$. As the $\mathrm{d}\overline{R}/\mathrm{d}E_{\textrm{dep}}(E_{\textrm{dep}})$ function in Eq.~\ref{eq:H1_extended} is normalized, $H^{(1)}(E_{\textrm{ph}})$ is describing a PDF for single-hit events from a given source. We also consider so-called ``multi-hit" events, which are events generated from simultaneous particle interactions in the detector. In general, the PDF solutions for multi-hit events are found by recursively convolving the single-hit solution from Eq.~\ref{eq:H1_extended}. An example of constructing a multi-hit PDF solution is shown in Sec..~\ref{sssec:extended_model_laserevents}.

Up to this point, the detector response model has been described without considering the detector energy resolution $\sigma_{\textrm{res}}$. While $\sigma_{\textrm{res}}$ can be incorporated into the model in different ways, this work assumes that the energy resolution is constant over $E_{\textrm{ph}}$. Therefore the single-hit model including the energy resolution $H^{(1)}(E_{\textrm{ph}},\sigma_{\textrm{res}})$ can be expressed as:
\begin{equation}\label{eq:H1_extended_res}
H^{(1)}(E_{\textrm{ph}},\sigma_{\textrm{res}})= H^{(1)}(E_{\textrm{ph}}) \ast G(E_{\textrm{ph}}\, | \, \mu = 0, \sigma_{\textrm{res}}),
\end{equation}
where $G(E_{\textrm{ph}}\, | \, \mu = 0, \sigma_{\textrm{res}})$ is a Gaussian function with a mean $\mu = 0$ and width of $\sigma_{\textrm{res}}$. One could also consider, for example, an energy resolution that depends on \neh{}. In which case, a Gaussian function with a width equal to the energy resolution at each \eh{}-pair peak $\sigma^{(j)}_{\textrm{res}}$ would be convolved with the corresponding $F^{(j)}_{\textrm{type}}$ function.

\subsubsection{Photon-Calibration Events}\label{sssec:extended_model_laserevents}
A typical way to calibrate the energy of HVeV detectors is to use a photon source. Specifically in Refs.~\cite{HVeVR1,HVeVR2}, a laser source of optical photons was pointed at one of the detector surfaces. The laser was pulsed at some known frequency $f_{\gamma}$, and, depending on the laser intensity, produced an average number of photons per pulse $\lambda$ that were detected. The probability of a given number of photons per laser event is given by a Poisson distribution with a mean of $\lambda$. These photon-calibration events are therefore examples of multi-hit events, where the probability distribution must also account for the probability of the simultaneous absorption of multiple photons in a single event. The differential rate $\mathrm{d}R/\mathrm{d}E_{\textrm{ph}}(E_{\textrm{ph}})$ for photon-calibration events is then given by:
\begin{equation}\label{eq:dRdEph_photon}
\frac{\mathrm{d}R}{\mathrm{d}E_{\textrm{ph}}}(E_{\textrm{ph}}) = f_{\gamma} \sum_{l=0}^{L}\textrm{Pois}(l\, | \, \lambda)H^{(l)}(E_{\textrm{ph}}),
\end{equation}
where $H^{(l)}(E_{\textrm{ph}})$ corresponds to the NTL energy produced in an event with $l \leq L$ photons absorbed. $H^{(0)}(E_{\textrm{ph}})$ corresponds to events with no photons absorbed. Such events may be present in the calibration data if the detector trigger is synchronized with the laser pulses and $\lambda$ is small. In the simplest scenario, $H^{(0)}(E_{\textrm{ph}}) = \delta(E_{\textrm{ph}})$. However the zeroth-\eh{}-pair peak can also take a more complex form, like the modified Gaussian noise peak described in Ref.~\cite{Ponce2020}. $H^{(1)}(E_{\textrm{ph}})$ corresponds to events with one photon absorbed and is generally given by Eq.~\ref{eq:H1_extended}. $H^{(l)}(E_{\textrm{ph}})$ with $l>1$ is calculated recursively as $H^{(l)}(E_{\textrm{ph}}) = H^{(l-1)}(E_{\textrm{ph}}) \ast H^{(1)}(E_{\textrm{ph}})$.

For photon-calibration events, we assume that the photons are absorbed sufficiently close to a detector surface such that these events can effectively be modelled as surface events created along the $z=0$ or $z=1$ plane. Furthermore for a spectrum of photon energies $E_{\gamma}$, $\mathrm{d}\overline{R}/\mathrm{d}E_{\textrm{dep}} = \mathrm{d}\overline{R}/\mathrm{d}E_{\gamma}$. For an LED source, $\mathrm{d}\overline{R}/\mathrm{d}E_{\gamma}$ is the normalized energy spectrum of the LED photons. Yet for a laser source like in Refs.~\cite{HVeVR1,HVeVR2}, the photons are monoenergetic, and therefore $\mathrm{d}\overline{R}/\mathrm{d}E_{\textrm{dep}} = \delta (E_{\textrm{dep}} - E_{\gamma})$. Moreover the laser used for the calibration in Refs.~\cite{HVeVR1,HVeVR2} produced 1.95~eV photons which, for silicon, always ionize exactly one~\eh{} pair per absorbed photon~\cite{Ramanathan2020}. Photons of this energy have an absorption length in silicon of $\mathcal{O}(10~\upmu\textrm{m})$ which, for a detector that is 4~mm thick~\cite{HVeVR2}, is sufficiently small to model these events as surface events. For this particular case, Eq.~\ref{eq:H1_extended} reduces to:
\begin{equation}\label{eq:H1_photoncal}
H^{(1)}(E_{\textrm{ph}}) = \frac{F^{(1)}_{\textrm{surf}}\left (\frac{E_{\textrm{ph}} - E_{\gamma}}{eV_{\textrm{bias}}} \right )}{eV_{\textrm{bias}}}.
\end{equation}

Again if we assume a constant energy resolution $\sigma_{\textrm{res}}$, Eq.~\ref{eq:dRdEph_photon} is convolved with a Gaussian function with a mean $\mu = 0$ and a width of $\sigma_{\textrm{res}}$ in order to compute $\mathrm{d}R/\mathrm{d}E_{\textrm{ph}}(E_{\textrm{ph}})$ with the energy resolution. The distinction between single-hit and multi-hit events displayed here is subtle yet important. For the case of low-energy photon-calibration events, the multi-\eh{}-pair peaks observed are not due to multiple \eh{} pairs ionized from a single absorbed photon, but rather simultaneously absorbed photons that each ionize a single \eh{} pair.

\subsubsection{Dark Matter Events}\label{sssec:extended_model_dmevents}
For any dark matter search experiment, a detector response model is required to determine the expected signal distribution of a DM candidate in the detector. Therefore Eq.~\ref{eq:H1_extended} can also be used to compute expected DM signals in HVeV detectors. Unlike photon-calibration events, DM interactions are considered to be single-hit events; generally DM models exclude the possibility of the simultaneous interaction of multiple DM particles with a detector. While the exact signal distribution will depend on the specific DM candidate that is modelled, we will look at examples of two DM candidates commonly searched for using HVeV detectors.

The first candidate is the dark photon that is modelled, for example, in Ref.~\cite{Hochberg2016}. In this model, non-relativistic dark photons with a mass $m_{A^{\prime}}$ constitute all relic dark matter. The interaction rate of dark photon absorption $R_{A^{\prime}}(m_{A^{\prime}},\varepsilon)$ depends on its mass and is proportional to the kinetic mixing parameter $\varepsilon$ that couples dark photons to standard model photons. In this model dark photons provide a monoenergetic source of energy deposition equal to its mass such that $\mathrm{d}\overline{R}/\mathrm{d}E_{\textrm{dep}} = \delta(E_{\textrm{dep}} - m_{A^{\prime}}c^{2})$, where $c$ is the speed of light. Substituting this into Eq.~\ref{eq:H1_extended} and noting that DM interactions are modelled as bulk-\eh{}-pair events, the differential rate of dark photon absorption $\mathrm{d}R_{A^{\prime}}/\mathrm{d}E_{\textrm{ph}}(E_{\textrm{ph}})$ is given by:
\begin{align}\label{eq:dRdEph_dpa}
\begin{split}
\frac{\mathrm{d}R_{A^{\prime}}}{\mathrm{d}E_{\textrm{ph}}}(E_{\textrm{ph}}) &= R_{A^{\prime}}(m_{A^{\prime}},\varepsilon) H^{(1)}(E_{\textrm{ph}}) \\
&= R_{A^{\prime}}(m_{A^{\prime}},\varepsilon) \Biggl( \sum_{j=1}^{J} p_{\textrm{eh}}(j\, | \, m_{A^{\prime}}c^{2}) \times \\
&\qquad \frac{1}{eV_{\textrm{bias}}}F^{(j)}_{\textrm{bulk-eh}} \left (\frac{E_{\textrm{ph}} - m_{A^{\prime}}c^{2}}{eV_{\textrm{bias}}} \right ) \Biggr).
\end{split}
\end{align}

The second candidate we consider is light DM that elastically scatters off of electrons, as described in Ref.~\cite{Essig2016}. In this model, the dark matter particle $\chi$ with mass $m_{\chi}$ is also assumed to constitute all relic DM, and scattering interactions with electrons are mediated via a dark-sector gauge boson. The total rate of DM-electron scattering interactions $R_{\chi}(m_{\chi},\bar{\sigma}_{e})$ is dependent on the DM mass as well as the effective DM-electron scattering cross section $\bar{\sigma}_{e}$. However, this DM-electron scattering process produces a spectrum of recoil energies $E_{r}$. Specifically in Ref.~\cite{Essig2016}, the recoil spectra are provided as rates over discrete recoil energy bins. Therefore the integral of $\mathrm{d}\overline{R}/\mathrm{d}E_{\textrm{dep}}$ in Eq.~\ref{eq:H1_extended} is replaced by a sum over weights $w_{k}$ corresponding to the recoil energies $E^{(k)}_{r}$, where the weights are normalized such that $\sum_{k} w_{k} = 1$. The differential rate of DM-electron scattering $\mathrm{d}R_{\chi}/\mathrm{d}E_{\textrm{ph}}(E_{\textrm{ph}})$ is then given by:
\begin{align}\label{eq:dRdEph_dme}
\begin{split}
\frac{\mathrm{d}R_{\chi}}{\mathrm{d}E_{\textrm{ph}}}(E_{\textrm{ph}}) &= R_{\chi}(m_{\chi},\bar{\sigma}_{e}) H^{(1)}(E_{\textrm{ph}}) \\
&= R_{\chi}(m_{\chi},\bar{\sigma}_{e}) \Biggl( \sum_{j=1}^{J} \sum_{k} p_{\textrm{eh}}(j\, | \, E^{(k)}_{r}) \times \\
&\qquad \frac{w_{k}}{eV_{\textrm{bias}}} \cdot F^{(j)}_{\textrm{bulk-eh}} \left (\frac{E_{\textrm{ph}} - E^{(k)}_{r}}{eV_{\textrm{bias}}} \right ) \Biggr).
\end{split}
\end{align}

The different rate functions in Eqs.~\ref{eq:dRdEph_dpa} and~\ref{eq:dRdEph_dme} do not yet include the detector energy resolution. As before we assume that $\sigma_{\textrm{res}}$ is constant over $E_{\textrm{ph}}$, and therefore the energy resolution is incorporated by convolving Eqs.~\ref{eq:dRdEph_dpa} and~\ref{eq:dRdEph_dme} with a Gaussian function with a mean $\mu = 0$ and a width of $\sigma_{\textrm{res}}$.

\subsection{Non-ionizing Energy Deposition}\label{ssec:extended_model_surfacetrapping}
\begin{figure}[t!]
\begin{center}
\includegraphics[width=0.95\columnwidth]{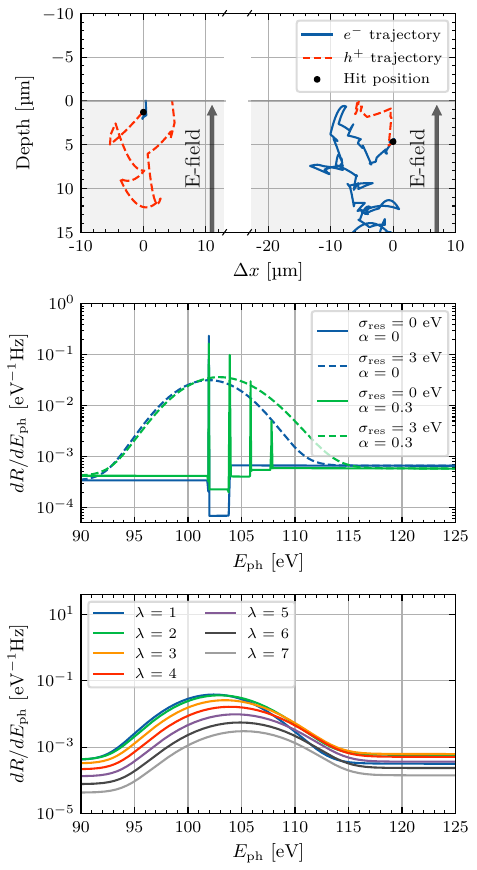}
\end{center}
\caption[Illustrations and examples of the surface trapping effect]{(top): Illustration of the hypothesized surface trapping effect as observed from simulation data using G4CMP~\cite{g4cmp_paper, g4cmp_code}. Two examples are shown of the trajectory of an ionized \eh{} pair in terms of the depth below the detector surface and the perpendicular $x$-coordinate relative to the hit position of the absorbed photon. The right example shows a typical event, where the electron eventually travels in the direction opposing the electric field. The left example shows a surface-trapped event, where the electron recombines with the detector surface before it can turn around. (middle and bottom): Examples of modelling the surface trapping effect using Eq.~\ref{eq:H1_photoncal_surftrap} and~\ref{eq:dRdEph_photon} with $V_{\textrm{bias}} = 100$\,V, $E_{\gamma} = 1.95$\,eV, $f_{\gamma} = 1$\,Hz, and arbitrary CT and II probabilities. The additional spikes seen in the middle plot demonstrate the contribution of non-ionizing energy deposition when $\alpha > 0$ which, when smeared by the energy resolution, widen and shift the \eh{}-pair peaks. This effect also causes a peak position dependence on $\lambda$, as demonstrated by the bottom plot with $\alpha = 0.3$.}
\label{fig:surface_trapping_examples}
\end{figure}

The detector response model can be extended further by modelling other phenomena that are observed in the detector. One such phenomenon is the apparent deposition of non-ionizing energy measured together with photon-calibration events. We surmise that this detector response effect occurs in HVeV detectors because of the observed dependence of the \eh{}-pair peak positions on $\lambda$, the average number of photons per laser or LED pulse~\cite{HONG2020}. One hypothesis is that some proportion of photons is absorbed directly into the aluminum fins of the phonon sensors. Another hypothesis is that, due to the random initial trajectory of electrons and holes after ionizing, there is some probability that both charges will happen to recombine at the nearest detector surface. This so-called surface trapping effect has been observed in detector simulations using G4CMP~\cite{g4cmp_paper, g4cmp_code}, and is illustrated in the top plot of Fig.~\ref{fig:surface_trapping_examples}. In any case, these hypotheses suppose that some proportion of photons will deposit some non-ionizing energy without generating a typical \eh{} pair that undergoes the bulk CT and II processes.

In the case of the hypothesized surface trapping effect, we can include this effect in the model by modifying the single-hit PDF for photon-calibration events described by Eq.~\ref{eq:H1_photoncal}. Let $\alpha$ be the probability of the created \eh{} pair to undergo surface trapping, where $0 \leq \alpha \leq 1$. That means there is a $(1-\alpha)$ probability that the \eh{} pair will propagate through the crystal, undergoing the typical bulk CT and II processes. For photons that undergo surface trapping, the deposited energy will only be the absorption energy of the photon $E_{\gamma}$. We are then able to include the surface trapping effect by modifying Eq.~\ref{eq:H1_photoncal} in the following way:

\begin{align}\label{eq:H1_photoncal_surftrap}
\begin{split}
H^{(1)}(E_{\textrm{ph}}) &\rightarrow \alpha \delta(E_{\textrm{ph}} - E_{\gamma}) + (1 - \alpha) H^{(1)}(E_{\textrm{ph}}) \\
&= \alpha \delta(E_{\textrm{ph}} - E_{\gamma}) \\
&\qquad \qquad + \frac{(1 - \alpha)}{eV_{\textrm{bias}}} F^{(1)}_{\textrm{surf}}\left (\frac{E_{\textrm{ph}} - E_{\gamma}}{eV_{\textrm{bias}}} \right ).
\end{split}
\end{align}

The multi-hit solution for photon-calibration data with the surface trapping effect is given by Eq.~\ref{eq:dRdEph_photon}, where $H^{(l)}(E_{\textrm{ph}})$ is found by recursively convolving Eq.~\ref{eq:H1_photoncal_surftrap} with itself. The result of including the surface trapping effect in the detector response model is illustrated in the bottom two plots of Fig.~\ref{fig:surface_trapping_examples}. Due to the presence of non-ionizing photons, each peak in the spectrum splits into multiple sub-peaks separated by $E_{\gamma}$, as seen in the middle plot of Fig.~\ref{fig:surface_trapping_examples} before resolution smearing. Each sub-peak corresponds to $q$ ionizing photons and $p$ non-ionizing photons, with the sub-peak location defined as $q\cdot eV_{\textrm{bias}} + (q+p)\cdot E_{\gamma}$. Note that a sub-peak corresponding to $p$ and $q$ is part of the function $H^{(q+p)}(E_{\textrm{ph}})$, rather than of the function $H^{(q)}(E_{\textrm{ph}})$. Therefore, in order to properly model the sub-structure of the $q\textrm{th}$ \eh{}-pair peak, modelling of higher \eh{}-pair peaks is required. When normalized, the underlying amplitudes of the sub-peaks in each \eh{}-pair peak follow a Poisson distribution of the number of non-ionizing photons with a mean of $\alpha \cdot \lambda$ (see Appendix~\ref{app:sub_peak} for more details). To include all the significant sub-peaks, it is recommended to set the maximum number of modelled peaks ($L$ in Eq.~\ref{eq:dRdEph_photon}) to a number exceeding the number of peaks in the region of interest by the mean number of non-ionizing photons plus at least 3 standard deviations of its distribution, i.e. by $(\alpha \cdot \lambda) + 3\cdot \sqrt{(\alpha \cdot \lambda)}$.

When the energy resolution is applied, the peak substructure from non-ionizing photons gets smeared and appears as a shift and a widening of the \eh{}-pair peaks, as shown in the bottom two plots of Fig.~\ref{fig:surface_trapping_examples}. When $\lambda$ increases, there is a greater contribution from non-ionizing photons, resulting in wider peaks that are shifted by a larger amount.

\section{\label{sec:results}Results}
\begin{figure}[t!]
\begin{center}
\includegraphics[width=1.0\columnwidth]{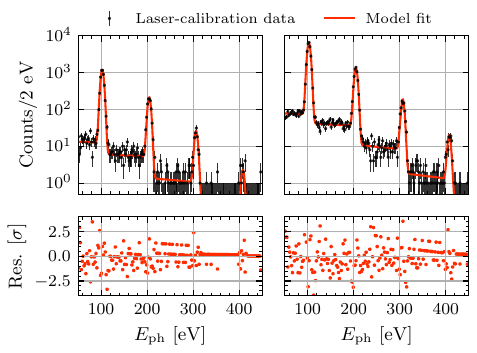}
\end{center}
\caption[Model fit to laser-calibration datasets.]{Results of fitting the exponential CTII and extended detector response model to two laser-calibration datasets from Ref.~\cite{HVeVR2} acquired with $V_{\textrm{bias}} = 100$\,V and $E_{\gamma} = 1.95$\,eV. The residuals from the fit results are shown in the bottom plots.}
\label{fig:laser_cal_data_fit}
\end{figure}

To demonstrate the performance of the exponential CTII and extended detector response model described in Secs.~\ref{sec:exp_model} and~\ref{sec:det_response}, we fit the model to laser-calibration data acquired from Ref.~\cite{HVeVR2}. Specifically, the model for photon-calibration events described by Eqs.~\ref{eq:dRdEph_photon} and \ref{eq:H1_photoncal} is fit to laser-calibration datasets from Ref.~\cite{HVeVR2} acquired with $V_{\textrm{bias}} = 100$\,V and $E_{\gamma} = 1.95$\,eV. The individual datasets differ by the laser intensity used during data acquisition, and thus by the value of $\lambda$. The fits of the model to two of these datasets are shown in Fig.~\ref{fig:laser_cal_data_fit}. For simplicity, we reduced the number of parameters in the fits by requiring:
\begin{equation}\label{eq:param_simplify}
\begin{split}
f_{\textrm{CTe}} = f_{\textrm{CTh}} &\equiv f_{\textrm{CT}}, \\
f_{\textrm{IIee}} = f_{\textrm{IIeh}} = f_{\textrm{IIhe}} =  f_{\textrm{IIhh}} &\equiv  \frac{f_{\textrm{II}}}{2}.
\end{split}
\end{equation}

We obtain the best-fit results from the fit to the left (right) laser-calibration dataset in Fig.~\ref{fig:laser_cal_data_fit} of $\lambda = 0.41 \,\pm \,0.01$ ($0.475 \,\pm \,0.005$), $\sigma_{\textrm{res}} = 3.30 \, \pm \,0.04$\,eV ($3.37 \,\pm \,0.02$\,eV), $f_{\textrm{CT}} = 11.6 \,\pm \,0.6$\,\% ($12.1 \,\pm 0\,.3$\,\%), and $f_{\textrm{II}} = 0.7 \,\pm 0\,.4$\,\% ($0.9 \pm \,0.2$\,\%). Within uncertainties, these results are consistent with the results obtained by fitting the flat CTII model from Ref.~\cite{Ponce2020} to the same datasets. The consistency of the results is expected, as the flat CTII model has previously demonstrated that it can accurately describe photon-calibration data~\cite{Ponce2020_2}. Figure~\ref{fig:laser_cal_data_fit} shows that the exponential CTII model is able to obtain equivalent results for this relatively simple scenario. However as will be shown below, the advantages of the model presented in this work become apparent when including additional detector response effects or when modelling different event types.

We can further evaluate the extended detector response model by performing a simultaneous fit of the model to multiple photon-calibration datasets. To do this, the simultaneous fit to multiple datasets is done separately for data acquired from two different experiments. The first are three laser-calibration datasets acquired in Ref.~\cite{HVeVR2}. The second are three LED-calibration datasets acquired at the Northwestern EXperimental Underground Site (NEXUS) at Fermilab (Batavia, IL). This NEXUS facility is located in the NUMI tunnel, which provides an overburden of 225\,mwe~\cite{Adamson:2015}, and hosts a Cryoconcept dry dilution refrigerator. The LED-calibration data reported in this work were acquired by one of four 1-cm-side HVeV detectors that were operated at NEXUS between May 14$^{\mathrm{th}}$ and July 27$^{\mathrm{th}}$, 2022. More information about the experiment design, data acquisition, and data analysis can be found in Ref.~\cite{HVeVR4}.

There are several key similarities and differences between laser-calibration datasets acquired in Ref.~\cite{HVeVR2} and the LED-calibration datasets acquired at the NEXUS facility reported in this work. In both cases, the data were acquired using an HVeV detector with an ``NF-C" sensor design~\cite{Ren:2021}. Both devices are constituted by a 10 $\times$ 10 $\times$ 4 mm$^3$ silicon target with two channels of Quasiparticle-trap-assisted Electrothermal-feedback Transition-Edge Sensors~\cite{Irwin:1995} (QETs) patterned on the top surface to measure the phonon signal. While the HVeV detector used to acquire the NEXUS data is not the same as the one used in Ref.~\cite{HVeVR2}, the substrate from both detectors belongs to the same silicon wafer. This means that the impurity levels in both detectors are likely to be similar.

Furthermore, both detectors generated an electric field throughout the bulk of the crystal by applying a high voltage to an aluminum electrode deposited on the detector surface opposite the surface patterned with the QETs; the QET surface of the detectors was kept grounded. In Ref.~\cite{HVeVR2}, laser-calibration data were acquired by emitting 1.95\,eV photons from a laser onto the center of the QET face of the detector. In contrast, the LED-calibration data from the NEXUS facility were acquired using an $\sim 2$\,eV LED collimated on the center of the electrode face of the detector. Yet the data from both detectors were acquired with $V_{\textrm{bias}} = + 100$\,V. This means that for the laser-calibration data from Ref.~\cite{HVeVR2}, the initial propagating charges are \textit{electrons}, whereas for the LED-calibration data from the NEXUS facility, the initial propagating charges are \textit{holes}. Moreover, because the LED-calibration data from the NEXUS facility illuminated the electrode face of the detector, any non-ionizing energy deposition due to photon absorption into the aluminum fins of the QETs is expected to be minimal, especially compared to the laser-calibration data from Ref.~\cite{HVeVR2}.

We fit the extended model assuming non-ionizing energy deposition caused by surface trapping (Eq.~\ref{eq:H1_photoncal_surftrap}) simultaneously to three laser-calibration datasets from Ref.~\cite{HVeVR2} and three LED-calibration datasets from the NEXUS facility, all acquired with $V_{\textrm{bias}} = 100$\,V. Each fit includes the parameters $\lambda_{1}$, $\lambda_{2}$, and $\lambda_{3}$ corresponding to the $\lambda$ value of each dataset, but includes only one value of $f_{\textrm{CT}}$, $f_{\textrm{II}}$, $\sigma_{\textrm{res}}$, and $\alpha$ for all datasets. For simplicity, we again reduced the number of parameters in the fit by imposing the requirements given by Eq.~\ref{eq:param_simplify}.

In the fit to the data from Ref.~\cite{HVeVR2}, we kept the energy of the laser photons fixed at $E_{\gamma}=1.95$\,eV, whereas in the fit to the NEXUS datasets, we allowed the energy of the LED photons to float. Furthermore, a measurement of the LED wavelength spectrum at 4\,K found the spread in photon energies to be $\sim 0.0012$\,eV, and therefore we can adequately treat the LED as a monoenergetic source of photons described by Eq.~\ref{eq:H1_photoncal}. For the NEXUS datasets, we additionally included parameters to calibrate the data. The calibration converts the pulse amplitude $A$ (in units of $\upmu\textrm{A}$) of each event to the total phonon energy $E_{\textrm{ph}}$. The fit includes three calibration parameters $c_{0}$, $c_{1}$, and $c_{2}$ that follow the equation:
\begin{equation} \label{eq:calibration}
A = c_{0} + c_{1}\cdot E_{\textrm{ph}} + c_{2}\cdot E_{\textrm{ph}}^{2},
\end{equation}
where the quadratic coefficient $c_{2}$ is included to account for any saturation effects in the QET sensors that can cause a non-linear response at higher energies~\cite{Ren:2021}.

\begin{figure}[t!]
\begin{center}
\includegraphics[width=1.0\columnwidth]{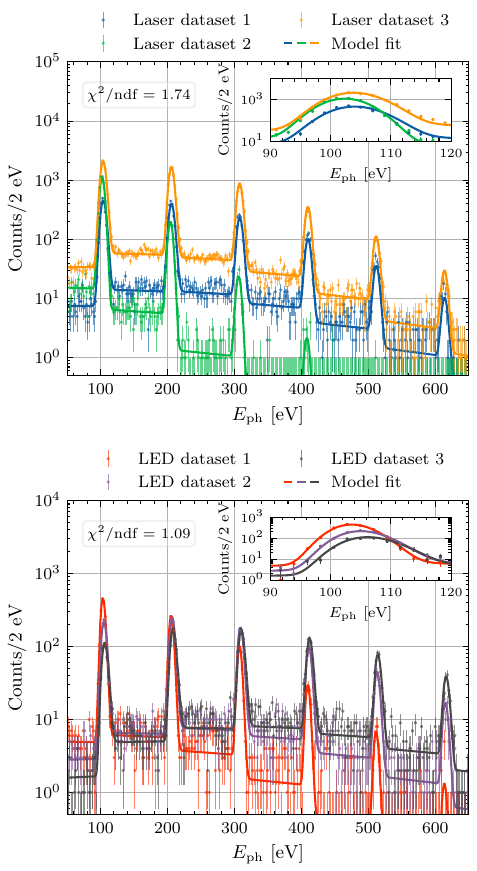}
\end{center}
\caption[Model fit with surface trapping to multiple laser-calibration datasets]{Results of simultaneous fits of the extended detector response model to three laser-calibration datasets from the HVeV detector in Ref.~\cite{HVeVR2} (top) and three LED datasets acquired from a HVeV detector at the NEXUS facility (bottom). All of the datasets were acquired with $V_{\textrm{bias}} = 100$\,V. The model assumes non-ionizing energy deposition caused by surface trapping as described by Eq.~\ref{eq:H1_photoncal_surftrap}. The inset plots show the data and fit enlarged around the first \eh{}-pair peak in order to clearly observe the peak shifts between datasets.}
\label{fig:laser_cal_multi_data_fit}
\end{figure}

\begin{table}[tb]
\caption{Best-fit results found by fitting the extended model assuming non-ionizing energy deposition caused by surface trapping (Eq.\ref{eq:H1_photoncal_surftrap}) simultaneously to multiple datasets. The results in the left column are determined from the fit to three laser-calibration datasets from Ref.~\cite{HVeVR2}, and the results in the right column are determined from the fit to three LED-calibration datasets from the NEXUS facility presented in this work. The fit to the NEXUS data also allowed the energy of the LED photons $E_{\gamma}$ to float, and included the calibration parameters $c_{0}$, $c_{1}$, and $c_{2}$.}

\begin{tabular}{ccc}
\toprule
 & Laser data (Ref.~\cite{HVeVR2}) & LED data (NEXUS) \\  \hline
 $\lambda_{1}$ & $3.18 \pm 0.04$ & $2.2 \pm 0.1$ \\
 $\lambda_{2}$ & $0.66 \pm 0.02$ & $4.2 \pm 0.2$ \\
 $\lambda_{3}$ & $2.90 \pm 0.03$ & $5.7 \pm 0.2$ \\
 $f_{\textrm{CT}}$ (\%) & $13.5 \pm 0.2$ & $11.5 \pm 0.2$ \\
 $f_{\textrm{II}}$ (\%) & $0.4 \pm 0.2$ & $0.0^{+0.3}_{-0.0}$\\ 
 $\sigma_{\textrm{res}}$ (eV) & $3.41 \pm 0.02$ & $2.71 \pm 0.06$ \\
 $\alpha$ (\%) & $36.9 \pm 0.7$ & $41 \pm 2$\\
 $E_{\gamma}$ (eV) & -- & $2.02 \pm 0.07$ \\
 $c_{0}$ ($\upmu\textrm{A}$) & -- & $(1.7 \pm 0.2)\times 10^{-3}$ \\
 $c_{1}$ ($\upmu\textrm{A}/\textrm{eV}$) & -- & $(9.102 \pm 0.006)\times 10^{-4}$ \\
 $c_{2}$ ($\upmu\textrm{A}/\textrm{eV}^{2}$) & -- & $(-242 \pm 7)\times 10^{-10}$ \\ \botrule
\end{tabular}

\label{tab:best_fit_vals}
\end{table}

The top and bottom plots of Fig.~\ref{fig:laser_cal_multi_data_fit} show the fit results to the datasets from Ref.~\cite{HVeVR2} and the dataset acquired at the NEXUS facility, respectively. The best-fit results of the fit parameters are listed in Tab.~\ref{tab:best_fit_vals}. As the CT and II probabilities of the initial propagating charge have the largest impact on the expected signal for photon-calibration events, we can interpret the values of $f_{\textrm{CT}}$ and $f_{\textrm{II}}$ from the fits to the datasets from Ref.~\cite{HVeVR2} and the datasets acquired at the NEXUS facility as the CT and II probabilities for electrons and holes, respectively. Therefore, these results suggest that for these detectors (that come from the same silicon wafer), the CT probability for electrons may be higher than for holes. By using Eq.~\ref{eq:p_tau} and knowing that the thickness of these detectors is 4\,mm, the fitted $f_{\mathrm{CT}}$ values in Tab.~\ref{tab:best_fit_vals} can be converted to the characteristic lengths of CT, giving $27.6 \pm 0.4$\,mm and  $32.7 \pm 0.6$\,mm for electrons and holes, respectively.

In both cases, the fit determined the amount of surface trapping to be $\sim 40$\,\%. Indeed the inset plots in Fig.~\ref{fig:laser_cal_multi_data_fit} that are enlarged around the one\,\eh{}-pair peak clearly show the peak position dependence on $\lambda$ --- a feature predicted when assuming surface trapping. Notably, the surface trapping probability found for the NEXUS data (where the LED photons are illuminated on the electrode face of the detector) is slighter higher compared to the data from Ref.~\cite{HVeVR2}. This strongly disfavours the hypothesis that non-ionizing energy deposition occurs due to photons being absorbed directly into the aluminum fins of the QETs. While this result supports the surface trapping hypothesis, we stress that it is just one interpretation of these data. Additional dedicated measurements are needed to confirm these detector response effects, as well as to understand the differences that these effects have on electrons and holes.

We can additionally demonstrate the differences in the detector response for expected DM signals which, as mentioned in Sec.~\ref{ssec:extended_model_singlemulti}, are modelled as bulk-\eh{}-pair events. To do this, we ran simulations using G4CMP~\cite{g4cmp_paper, g4cmp_code} of two DM models by generating events within the bulk of a silicon HVeV detector with $V_{\textrm{bias}} = 100$\,V. The first simulated signal is dark photon absorption following Ref.~\cite{Hochberg2016} with a dark photon mass $m_{A^{\prime}} = 10$\,eV$/c^{2}$ and kinetic mixing parameter $\varepsilon = 10^{-12}$, and the second is DM-electron scattering following Ref.~\cite{Essig2016} with a DM mass $m_{\chi} = 5$\,MeV$/c^{2}$, effective DM-electron scattering cross section $\bar{\sigma}_{e} = 10^{-33}$\,cm$^{2}$, and a DM form factor of $F_{\textrm{DM}} = 1$. The total number of events in both simulations were determined assuming an exposure of 6\,gram-days, and the ionization PMFs in the simulations are computed using the binomial approach taken in Refs.~\cite{HVeVR1,HVeVR2}. Finally, we assumed an energy resolution of $\sigma_{\textrm{res}} = 3$\,eV and the following CT and II probabilities: $f_{\mathrm{CTe}} = f_{\mathrm{CTh}} = 10$\,\% and $f_{\mathrm{IIee}} = f_{\mathrm{IIeh}} = f_{\mathrm{IIhe}} =f_{\mathrm{IIhh}} = 1$\,\%.

\begin{figure}[t!]
\begin{center}
\includegraphics[width=1.0\columnwidth]{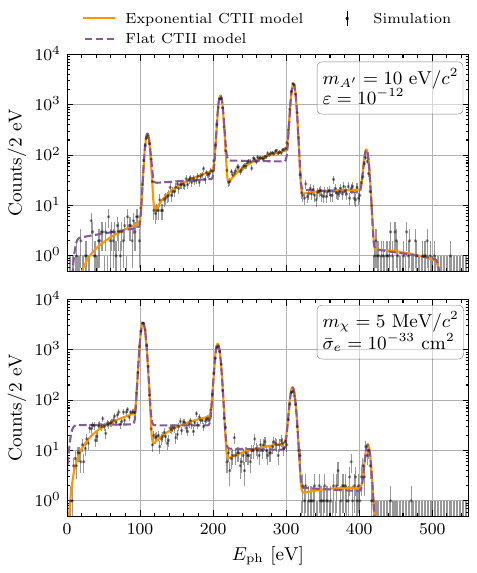}
\end{center}
\caption[Expected DM signals in an HVeV detector using G4CMP simulations]{Simulated data of events generated in the bulk of a silicon HVeV detector with $V_{\textrm{bias}} = 100$\,V for DM signals using G4CMP~\cite{g4cmp_paper, g4cmp_code}. The simulations were run for (top) dark photon absorption following Ref.~\cite{Hochberg2016} with $m_{A^{\prime}} = 10$\,eV$/c^{2}$ and $\varepsilon = 10^{-12}$ and (bottom) DM-electron scattering following Ref.~\cite{Essig2016} with $m_{\chi} = 5$\,MeV$/c^{2}$, $\bar{\sigma}_{e} = 10^{-33}$\,cm$^{2}$, and $F_{\textrm{DM}} = 1$. For running the simulations, we assumed an exposure of 6\,gram-days, an energy resolution of $\sigma_{\textrm{res}} = 3$\,eV and CT and II probabilities of $f_{\mathrm{CTe}} = f_{\mathrm{CTh}} = 10$\,\% and $f_{\mathrm{IIee}} = f_{\mathrm{IIeh}} = f_{\mathrm{IIhe}} =f_{\mathrm{IIhh}} = 1$\,\%. The solid, orange curves are the expected signals computed using our model following Eqs.~\ref{eq:dRdEph_dpa} and~\ref{eq:dRdEph_dme}. For comparison, the dashed, purple curves are the expected signals computed using the flat CTII model from Ref.~\cite{Ponce2020}.}
\label{fig:dm_sim}
\end{figure}

The results of these simulations are shown in Fig.~\ref{fig:dm_sim}, which also shows the expected signal for the two DM models computed using the exponential CTII and extended detector response model from this work following Eqs.~\ref{eq:dRdEph_dpa} and~\ref{eq:dRdEph_dme}. It is important to note that G4CMP-based simulations model CT and II processes using the same PDF described by Eq.~\ref{eq:Pi_z} and are parametrized using characteristic lengths~\cite{g4cmp_paper}. The consistency between the solid, orange curves and simulated data shown in Fig.~\ref{fig:dm_sim} therefore provides a verification of the analytical solutions found for the various CT and II processes modelled. For comparison, Fig.~\ref{fig:dm_sim} also shows the expected DM signals computed using the flat CTII model from Ref.\cite{Ponce2020} which, unlike the exponential CTII model, does not distinguish between surface and bulk event types. Differences in the energy spectra for different event sources are evident by comparing Fig.~\ref{fig:dm_sim} with the energy spectra seen in Figs.~\ref{fig:laser_cal_data_fit} and~\ref{fig:laser_cal_multi_data_fit}. Our modelling expects the signal shape in the between-peak regions to differ for a bulk-\eh{}-pair source of events, such as DM, compared to surface events, such as photon-calibration data. The signal shape in the between-peak regions has a lot more curvature for bulk-\eh{}-pair events, in contrast to the relatively straight signal shape between the peaks for surface events. This feature is not captured by the flat CTII model, as evident in Fig.~\ref{fig:dm_sim}.

\section{\label{sec:discussion}Discussion}
The exponential CTII model introduced in Sec.~\ref{sec:exp_model} addresses several limitations of the flat CTII model from Ref.~\cite{Ponce2020}. This more robust model adopts a more physically motivated approach to describe CT and II processes in phonon-based crystal detectors. Consequently, it effectively characterizes the detector response across a range of event types and allows for the differentiation of CT and II probabilities for electrons and holes. These advantages, coupled with the expanded detector response model detailed in Sec.~\ref{sec:det_response}, provide a more accurate representation of the detector's response to different sources of events. 

The exponential CTII model combined with the extended response model assuming non-ionizing energy deposition caused by surface trapping is shown in Fig.~\ref{fig:laser_cal_multi_data_fit} to provide an accurate description of laser-calibration data from Ref.~\cite{HVeVR2} and LED-calibration data acquired from the NEXUS facility reported in this work. While these results are encouraging, we note that this model may not encompass all relevant detector response effects that may occur. Rather, these results motivate us to acquire future measurements in order to continue investigating the full extent of these effects. Apart from understanding detector response effects, the result of the fit to the NEXUS data shown in Fig.~\ref{fig:laser_cal_multi_data_fit} and Tab.~\ref{tab:best_fit_vals} additionally demonstrates how the extended detector response model can be utilized to calibrate the detector. Incorporating these additional detector response effects can help to improve the accuracy of the energy calibration. 

Furthermore, Fig.~\ref{fig:dm_sim} illustrates how the signal shape differs for DM models, where the detector response is instead modelled by bulk-\eh{}-pair events. Indeed the analytical solutions to the exponential CTII model presented in this work match the spectra of events simulated using G4CMP, whereby CT and II processes are also parametrized using characteristic lengths~\cite{g4cmp_paper}. The ability of the exponential CTII model to describe different event types provides a large advantage over the flat CTII model from Ref.~\cite{Ponce2020}, and can enable statistical discrimination between different background signals and expected DM signals.

Yet it is equally important to outline the limitations of the exponential CTII model. Physically, this model provides no description of the underlying mechanisms of CT or II processes, and simply assumes that there is some probability that these processes occur due to impurities throughout the crystal bulk. The characteristic lengths of these processes, $\tau_i$, could depend not only on the density of impurities, but also on the strength of the electric field, prebiasing history, ``baking'' history (impurity neutralization by detector irradiation) and temperature \cite{sundqvist, phipps}. Furthermore,
as mentioned in Sec.~\ref{sec:exp_model}, the exponential CTII model limits the analytical solutions to second-order processes for surface and bulk-single-charge events, and first-order processes for bulk-\eh{}-pair events. This limitation is a practical necessity, as including higher-order processes would exponentially increase the number and the complexity of solutions to solve for. However, the probability of II has been found in measurements to be of the order of 1\,\%~\cite{Ponce2020_2,HVeVR2}. Therefore second- and third-order processes are expected to be extremely subdominant with probabilities $\ll 0.01\,\%$. Nevertheless, this limitation is quantitatively assessed for each event type in Appendix~\ref{app:model_limit}. 

While the detector response effects described in this work can also be modelled using Monte Carlo (MC) simulations, the exponential CTII model also provides a considerable computational advantage. For instance, calculating a DM exclusion limit requires the calculation of the expected DM signal, which itself may need to be computed for a large selection of CT or II values. The analytical solutions found for the exponential CTII model provide a quick and easy method for generating multiple DM signals with different CT or II probabilities compared to the computationally intense method of running many MC simulations.

The bulk CT and II processes modelled in this work are also present in other solid-state DM experiments, including SENSEI and DAMIC-M~\cite{senseicollaboration2023sensei,DAMIC2023}. These experiments use charge-coupled devices (CCDs) that read out the amount of charge collected at each pixel. As such, charges that become trapped within the bulk of the crystal result in a signal loss. Furthermore, because charge packets are required to move from pixel to pixel, the drift length of charges is typically larger for CCD detectors compared to HVeV detectors. The proposed Oscura experiment aims to account for this signal loss by considering that these trapped charges may be released at a later time and measured as single-electron events~\cite{Cervantes-Vergara_2023}. This bulk trapping is distinct from charge transfer inefficiency that occurs in CCD detectors, whereby charges are lost to surrounding pixels as the charge packet is moved from pixel to pixel. As shown in this work, the advantage of using phonon-based crystal detectors, including HVeV detectors, lies in the ability to exploit the non-quantized \eh{}-pair peaks regions on the energy spectrum to extract CT and II parameters as well as to differentiate among different sources of events.

Future plans involve dedicated detector response investigations using HVeV detectors. For example, taking CT and II measurements using crystals of different impurity levels while varying the voltage bias or the amount of prebiasing applied to the crystals can help develop our understanding of the factors that contribute to CT and II. Taking measurements with different voltage polarities will allow us to probe the differences in CT and II processes for electrons and holes. Furthermore, we aim to explore additional detector response effects and phenomena, including sources of non-ionizing energy deposition. While these proposed measurements can be made using a photon-calibration source, finding a source of low-energy, bulk-\eh{}-pair events would additionally allow us to investigate the detector response expected for DM signals. The nuclear-recoil ionization yield measurement in Ref.~\cite{impact2023} demonstrates a method of producing such events by measuring low-energy neutron recoils off of the nuclei of an HVeV detector. The model presented in this work not only provides a more robust understanding of the detector response effects in phonon-based crystal detectors, but can be utilized to help discriminate among different sources of events in order to improve the sensitivity of DM search experiments.

\begin{acknowledgments}
We would like to thank Francisco Ponce and Chris Stanford for helpful initial discussions about understanding CT and II processes and how to model them. We would further like to thank Michael Kelsey for helpful discussions about G4CMP simulations and Elham~Azadbakht for contributions implementing CT and II processes in G4CMP. We also thank Francesco~Toschi and Sukeerthi~Dharani for comments on a draft version of this manuscript. Funding and support were received from the Deutsche Forschungsgemeinschaft under the Emmy Noether Grant No. 420484612, NSERC Canada, the Canada First Excellence Research  Fund, the Arthur B. McDonald Institute (Canada), the U.S. Department of Energy Office of High Energy Physics, and from the National Science Foundation (NSF). This work was supported in part under NSF Grant No. 2111324. Fermilab is operated by Fermi Research Alliance, LLC, under Contract No. DE-AC02-37407CH11359 with the U.S. Department of Energy, and computing support and resources were provided by SciNet (\cite{scinet}), the Digital Research Alliance of Canada (\cite{alliancecan}), and SLAC.
\end{acknowledgments}

\appendix

\section{Single-\eh{}-Pair Solutions}\label{app:single_eh_solution}
This appendix provides details on how to find the unique solutions for single-\eh{}-pair events. As mentioned in Sec.~\ref{sec:exp_model}, the events are categorized into three distinct classes: surface event, bulk-single-charge events, and bulk-\eh{}-pair events. Recall that both propagating electrons and holes have three processes which can occur: charge trapping (CTe and CTh), creation of a charge of the same kind (IIee and IIhh), and creation of a charge of the opposite kind (IIeh and IIhe). When writing the equation for the probability of a particular scenario occurring, we must consider the probabilities of all possible processes. For example, consider the probability of CT for a propagating electron as a function of $\Delta z$, $P_{\mathrm{CTe}}(\Delta z)$. The probability of CTe occurring at a distance $\Delta z$ travelled is the probability of CTe occurring at $\Delta z$ \textit{and} IIee not occurring by $\Delta z$ \textit{and} IIeh  not occurring by $\Delta z$. Using Eqs.~\ref{eq:Cbar_z} and~\ref{eq:Pi_z}, this scenario is described by:

\begin{equation}\label{eq:CTe_example}
\begin{split}
    P_{\mathrm{CTe}}(\Delta z) &=  \frac{1}{\tau_{\mathrm{CTe}}}e^{-\Delta z/\tau_{\mathrm{CTe}}} \cdot \overline{C}_{\mathrm{IIee}}(\Delta z) \cdot \overline{C}_{\mathrm{IIeh}}(\Delta z)\\
    &= \frac{1}{\tau_{\mathrm{CTe}}}e^{-\Delta z \left(1/\tau_{\mathrm{CTe}} + 1/\tau_{\mathrm{IIee}} + 1/\tau_{\mathrm{IIeh}} \right)} \\
    & =  \frac{1}{\tau_{\mathrm{CTe}}}e^{-\Delta z T_{\mathrm{e}}},
\end{split}
\end{equation}
where we define
\begin{equation} \label{eq:Te_def}
    T_{\mathrm{e}} \equiv \frac{1}{\tau_{\mathrm{CTe}}} + \frac{1}{\tau_{\mathrm{IIee}}} + \frac{1}{\tau_{\mathrm{IIeh}}}.
\end{equation}

Equivalent equations to Eq.~\ref{eq:CTe_example} can be found for the other five unique processes and by defining
\begin{equation} \label{eq:Th_def}
    T_{\mathrm{h}} \equiv \frac{1}{\tau_{\mathrm{CTh}}} + \frac{1}{\tau_{\mathrm{IIhe}}} + \frac{1}{\tau_{\mathrm{IIhh}}}.
\end{equation}

We also need to consider the probability of a charge propagating a distance $\Delta z$ without any CT or II process occurring. In this context, it is evaluating the probability of a charge reaching one of the crystal surfaces after travelling a distance $\Delta z$. We denote this probability as $P_{\mathrm{S}}(\Delta z)$. For a propagating electron, this probability is given by:

\begin{equation} \label{eq:prob_surface_example}
    \begin{split}
        P_{\mathrm{S}}(\Delta z) &=\overline{C}_{\mathrm{CTe}}(\Delta z) \cdot \overline{C}_{\mathrm{IIee}}(\Delta z) \cdot \overline{C}_{\mathrm{IIeh}}(\Delta z) \\
        &= e^{-\Delta z T_{\mathrm{e}}}.
    \end{split}
\end{equation}

Equivalently, the probability for a propagating hole to reach a surface without a CT or II process occurring as a function of $\Delta z$ is $P_{\mathrm{S}}(\Delta z) = e^{-\Delta z T_{\mathrm{h}}}$. Equation~\ref{eq:prob_surface_example} says that if a charge travels the full length of the detector (i.e. $\Delta z=1$), the probability that it does not undergo a CT or II process is $e^{- T_{\mathrm{e/h}}}$.

So far, we have only described probabilities of propagating charges undergoing CT and II processes as a function of the distance travelled. However, detectors do not directly measure the distance charges are able to travel. As mentioned in Sec.~\ref{sec:exp_model}, the NTL energy measured by the detector is proportional to the distance travelled by the charges. In this parametrization, the energy measured due to ionization is therefore equal to the total distance travelled by all propagating charges involved with an event. The probability of measuring some energy $E$ (where $E$ is in \neh{}-energy space, denoted as $E_{\textrm{neh}}$ is Sec.~\ref{sec:exp_model}) can be described as the sum of probabilities whereby the total distance travelled by all charges $z_{\mathrm{tot}}$ is equal to $E$. While there is no direct constraint on $z_{\mathrm{tot}}$, and thus $E$, individual charges are constrained by the bounds of the crystal surfaces (i.e. $0 \leq z \leq 1$).

Using the understanding of how the measured energy is related to the distance travelled by the propagating charges, we can start by solving for the almost trivial solutions, which will also create a set of base equations in which all other solutions can be found. We consider the solutions for charges propagating toward the $z=0$ and $z=1$ surfaces separately. For a charge travelling toward the $z=1$ surface that starts at a position $z_{0}$, the total energy that can be measured by the charge is $E = 1 - z_{0}$. Conversely for a charge travelling toward the $z=0$ surface that starts at a position $z_{0}$, the total energy that can be measured by the charge is $E=z_{0}$. Using Eq.~\ref{eq:prob_surface_example}, we can write the probability distribution for a charge reaching the $z=0$ and $z=1$ surfaces as:
\begin{equation} \label{eq:base_solution_1}
    \begin{split}
        P^{0}_{\mathrm{S},\,q}(E,z_{0}) &= \delta \left(E - z_{0} \right)e^{-z_{0}T_{q}} \\
        P^{1}_{\mathrm{S},\,q}(E,z_{0}) &= \delta \left(E - 1 + z_{0} \right)e^{-(1-z_{0})T_{q}} ,
    \end{split}
\end{equation}
where the subscript $q = (\mathrm{e},\, \mathrm{h})$ indicates if the charge is an electron or hole, and the superscripts $0$ and $1$ indicate which surface the charge is travelling toward. Next, we consider the probability distribution of measuring an energy $E$ \textit{before} some CT or II process occurs. For now, whether the process is CT or II does not matter. These base solutions are found using Eq.~\ref{eq:CTe_example} as a framework and solved separately for charges propagating toward the $z=0$ and $z=1$ surfaces. For charges that undergo some process $i$ and start at a position $z_{0}$, these probability distributions $P_{i}(E,z_{0})$ go as:

\begin{align} \label{eq:base_solution_2}
\begin{split}
 P^{0}_{i}(E,z_{0}) &= 
 \begin{cases}
\frac{1}{\tau_{i}}e^{-E\cdot T_{q}} & 0 \leq E < z_{0}\\
0 & \mathrm{else},
 \end{cases} \\
 P^{1}_{i}(E,z_{0}) &= 
 \begin{cases}
\frac{1}{\tau_{i}}e^{-E\cdot T_{q}} & 0 \leq E < 1- z_{0}\\
0 & \mathrm{else},
 \end{cases}
 \end{split}
 \end{align}
where again the superscripts 0 and 1 indicate the direction of propagation.

Some specific solutions can immediately be found from Eq.~\ref{eq:base_solution_2}. Specifically when a charge undergoes CT, the charge can no longer propagate and produce more energy or undergo additional processes. Therefore Eq.~\ref{eq:base_solution_2} are also the solutions for the probability distribution of a CT process for a charge starting at a position $z_{0}$. This is decisively not the case for a charge that undergoes II, as both the initial charge and the newly created charge will continue to traverse the crystal and increase the energy measured. Nevertheless, Eqs.~\ref{eq:base_solution_1} and~\ref{eq:base_solution_2} provide the necessary foundation for calculating the probability distribution for any specific scenario. Equipped with the base equations of the probability distributions in energy for propagating charges travelling in either direction with some starting position, we can now analytically solve for the single-\eh{}-pair probability distributions corresponding to specific events and scenarios.

\subsection{\label{sssec:surface_Events}Surface Events}
A surface event starts with the creation of either an \e{} or a \h{} at the starting position $z_{0} = 0$ or $z_{0}=1$. Knowing the polarity of the voltage bias and the starting position will necessarily decide whether the propagating charge is an \e{} or a \h{}. We can reduce the number of solutions to solve for by accounting for the symmetries that exist in this scenario. First, if a solution is found for when an \e{} is the initial propagating charge, the solution for when a \h{} is the initial propagating charge can be immediately found by swapping the $\tau$ and $T$ parameters. Therefore we need only to keep the distinction between charges that are the same as the initial charge and charges that are the opposite. This is done by replacing the ``e" and ``h" labels in the $\tau$ and $T$ parameters with ``s" and ``o" labels to indicate the same or opposite charge. Second, the solutions should be the same whether the charge is propagating toward the $z = 0$ or $z=1$ surface. Therefore solutions need only be found for one direction of propagation. However for good practice, the solutions were solved for both directions of propagation and are confirmed to give matching results. Here we will solve some of the solutions for when the initial charge is propagating toward $z=1$. 

We start by defining the probability distribution of the initial starting position of the charges. For surface charges that propagate toward $z=1$, the probability distribution of the initial starting position $P_{\mathrm{surf}}(z_0)$ trivially goes as:

\begin{equation}\label{eq:Pdist_surf}
    P_{\mathrm{surf}}(z_0) = \delta \left( z_{0} \right).
\end{equation}

Next, we can begin to solve for the solutions corresponding to specific scenarios. These probability distributions are indexed as $P_{k}(E)$, where $k$ iterates through the different solutions. The first, and easiest, solution to solve for is the case where the initial charge reaches the surface without a CT or II process occurring. This probability $P_{0}(E)$ is found by combining Eqs.~\ref{eq:base_solution_1} and~\ref{eq:Pdist_surf}.

\begin{equation}\label{eq:Psurf_0}
\begin{split}
    P_{0}(E) &= \int_{-\infty}^{\infty} P_{\mathrm{surf}}(z_0) P^{1}_{\mathrm{S},\,\mathrm{s}}(E,z_{0}) \mathrm{d}z_{0} \\
    &= \int_{-\infty}^{\infty} \delta \left( z_{0} \right) \delta \left(E - 1 + z_{0} \right)e^{-(1-z_{0})T_{s}} \mathrm{d}z_{0}\\
    &= \delta \left(E - 1 \right)e^{-T_{\mathrm{s}}}
\end{split}
\end{equation}

The solution to $P_{0}(E)$ is a delta function at $E=1$ with an amplitude of $e^{-T_{s}}$. This makes sense, as a charge travelling from $z=0$ to $z=1$ will produce exactly one \eh{} pair worth of energy. Next, we can solve for the solution when the initial charge undergoes CT, $P_{1}(E)$, which is found by combining Eqs.~\ref{eq:base_solution_2} and~\ref{eq:Pdist_surf},

\begin{align}\label{eq:Psurf_1}
\begin{split}
    P_{1}(E) &= \int_{-\infty}^{\infty} P_{\mathrm{surf}}(z_0) P^{1}_{\mathrm{CTs}}(E,z_{0}) \mathrm{d}z_{0}  \\
    &= \int_{-\infty}^{\infty} \delta \left( z_{0} \right) P^{1}_{\mathrm{CTs}}(E,z_{0}) \mathrm{d}z_{0}  \\
    &= P^{1}_{\mathrm{CTs}}(E,0) \\
    & = \begin{cases}
    \frac{1}{\tau_{\mathrm{CTs}}}e^{-T_{\mathrm{s}} \cdot E } & 0 \leq E < 1\\
    0 & \mathrm{else.}
    \end{cases} 
\end{split}
\end{align}

The next scenario to consider is the case where the initial charge creates a new, like charge, and both the original charge and the new charge happen to reach the surface. We call this probability distribution $P_{2}(E)$. Let $z_{\mathrm{II}}$ be the position where the new charge is created. It is important to remember that the quantity we are interested in is the total distance travelled by all the charges in the scenario. It is helpful to think about the energy gained by each ``segment" of charge propagation. In this scenario, there are three such segments: the initial charge that travels from $z_{0}$ to $z_{\mathrm{II}}$, and the initial and additional charges that each travel from $z_{\mathrm{II}}$ to $z=1$. It is additionally helpful to think of $E$ as fixed number that constrains the problem. We want to find the combinations of segments that result in $E$ total energy.

Let $E_{2a}$ and $E_{2b}$ be the energy contributions from each of the two charges after II. The combined energy contribution from both charges after II is therefore $E_{2} = E_{2a} + E_{2b}$. If $E_{1}$ is the energy contribution of the initial charge before II, then the total energy measured will be $E = E_{1} + E_{2}$. We also know that $E_{1}$ can be expressed as $z_{\mathrm{II}} - z_{0}$. Rearranging gives $z_{\mathrm{II}} = E - E_{2} + z_{0}$. What we have done is expressed $z_{\mathrm{II}}$ not as a position in the crystal, but rather in terms of energy contributions. The choice of expressing the parameters this way initially seems odd. After all, we already know that $E_{2a}$ and $E_{2b}$ are equal in this scenario. But that is not true for all scenarios, and it turns out that this way of formulating the problem provides a generic framework for solving all of the solutions.

We first need to find the probability that the energy contribution after II is $E_{2}$. This probability is equal to the probability that one charge contributes an energy of $E_{2a}$ times the probability that the other charge contributes an energy of $E_{2} - E_{2a}$, given a starting position of $z_{\mathrm{II}}$ and summed over all possibilities of $E_{2a}$. Expressed in terms of the base equations from Eq.~\ref{eq:base_solution_1} gives:

\begin{equation}\label{eq:Psurf_2_1}
\begin{split}
    P(E_{2}) &=   \int_{-\infty}^{\infty}\mathrm{d}E_{2a}\,  P^{1}_{\mathrm{S},\,\mathrm{s}}(E_{2a},E - E_{2} + z_{0}) \times \\ 
    &\qquad \qquad P^{1}_{\mathrm{S},\,\mathrm{s}}(E_{2} - E_{2a},E - E_{2} + z_{0}) \\
    &= \int_{-\infty}^{\infty}\mathrm{d}E_{2a}\, \delta \left(E_{2a} - 1 + E - E_{2} + z_{0} \right) \times \\
    & \qquad \qquad  \delta \left(E_{2} - E_{2a} - 1 + E - E_{2} + z_{0} \right) \times \\
    & \qquad \qquad e^{-2(1-E + E_{2} - z_{0})T_{\mathrm{s}}} \\
    &= \delta \left(2E - E_{2} - 2 + 2z_{0} \right) e^{-2(1-E + E_{2} - z_{0})T_{\mathrm{s}}}. \\
\end{split}
\end{equation}

Next, the probability of having a total energy of $E$ for a given starting position $z_{0}$ is the probability that the energy contribution after II is $E_{2}$ and the energy contribution before II is $E - E_{2}$. We find this by combining Eqs.~\ref{eq:Psurf_2_1} and~\ref{eq:base_solution_2}, and summing over all possibilities of $E_{2}$,

\begin{align}\label{eq:Psurf_2_2}
\begin{split}
    P(E,z_{0}) &=   \int_{-\infty}^{\infty}\mathrm{d}E_{2}\,  P(E_{2}) \times P^{1}_{\mathrm{IIss}}(E-E_{2},z_{0}) \\
    &=  \int_{E-1+z_{0}}^{E}\mathrm{d}E_{2}\, \delta \left(2E - E_{2} - 2 + 2z_{0} \right) \times \\
    & \qquad \frac{1}{\tau_{\mathrm{IIss}}} e^{-2(1-E + E_{2} - z_{0})T_{\mathrm{s}}} e^{-(E - E_{2})T_{\mathrm{s}}} \\
    &=  \int_{E-1+z_{0}}^{E}\mathrm{d}E_{2}\, \delta \left(2E - E_{2} - 2 + 2z_{0} \right) \times \\
    & \qquad \frac{1}{\tau_{\mathrm{IIss}}} e^{-(2-E + E_{2} - 2z_{0})T_{\mathrm{s}}}  \\
    &= \begin{cases}
    \frac{1}{\tau_{\mathrm{IIss}}} e^{-T_{\mathrm{s}} \cdot E} & 1 - z_{0} \leq E < 2(1-z_{0}) \\
    0 & \mathrm{else.}
    \end{cases}
\end{split}
\end{align}

The final step to find $P_{2}(E)$ is to integrate over all $z_{0}$,

\begin{align}\label{eq:Psurf_2_3}
\begin{split}
P_{2}(E) &= \int_{-\infty}^{\infty} P(E,z_{0}) P_{\mathrm{surf}}(z_0) \mathrm{d}z_{0} \\
    &= \begin{cases}
    \frac{1}{\tau_{\mathrm{IIss}}} e^{-T_{\mathrm{s}} \cdot E} & 1 \leq E < 2 \\
    0 & \mathrm{else.}
    \end{cases}
\end{split}
\end{align}

The next scenario to consider is when the initial charge creates an opposite charge, and both the original and new charge happen to reach the surface. We call this probability distribution $P_{3}(E)$. Like the previous scenario, the additional charge is created at a position $z_{\mathrm{II}}$, and we need to find the probabilities of each segment of charge propagation. Because of how we formulated the problem, the way to solve for $P_{3}(E)$ is exactly the same as how to solve for $P_{2}(E)$ with just two key substitutions. The II process created an opposite charge, and that opposite charge will propagate toward the $z=0$ surface. Therefore, one of the $P^{1}_{\mathrm{S},\,\mathrm{s}}$ terms in Eq.~\ref{eq:Psurf_2_1} is replaced with $P^{0}_{\mathrm{S},\,\mathrm{o}}$ with the same inputs. And because the process in this scenario is II to an opposite charge, the $P^{1}_{\mathrm{IIss}}$ term in Eq.~\ref{eq:Psurf_2_2} is replaced with $P^{1}_{\mathrm{IIso}}$, also with the same inputs. Making these substitutions and solving for $P_{3}(E)$ gives:

\begin{align}\label{eq:Psurf_3}
P_{3}(E) = \begin{cases}
    \frac{1}{\tau_{\mathrm{IIso}}} e^{T_{\mathrm{o}} - T_{\mathrm{s}} - T_{\mathrm{o}} \cdot E} & 1 \leq E < 2 \\
    0 & \mathrm{else}.
    \end{cases}
\end{align}

The solutions for other scenarios can be found by employing the same logic of considering segments of energy contribution, using the correct combination of base equations, and nested integrals. In total, there are 28 solutions found for surface charge events, all of which are catalogued in the Supplemental Material. The individual solutions are shown together in the top plot of Fig.~\ref{fig:1eh_solutions}.

\subsection{\label{sssec:bulk_singlecharge_Events}Bulk-Single-Charge Events}
A singe-charge bulk event starts with the creation of either an \e{} or \h{} at some starting position $z_{0}$ that ranges between $z=0$ and $z=1$. As with surface charges, we can use the same symmetry arguments to reduce the number of solutions to solve for. Again, the e and h labels in the subscripts are replaced with s and o to indicate charges that are the same and opposite as the initial charge, and solutions are only needed to be found for charges propagating in one direction. For bulk-single-charge events, the probability distribution of $z_{0}$ is defined as a uniform distribution between $z=0$ and $z=1$:

\begin{equation}\label{eq:Pdist_bulk}
P_{\mathrm{bulk}}(z_0) = 
\begin{cases}
     1 & 0 \leq z_{0} \leq 1 \\
     0 & \mathrm{else}.
\end{cases}
\end{equation}

We can again consider the simplest scenarios to solve for $P_{0}(E)$ (the charge reaches the surface),  $P_{1}(E)$ (the charge undergoes CT),  $P_{2}(E)$ (the charge undergoes II to the same charge and both charges reach the surface), and $P_{3}(E)$ (the charge undergoes II to the opposite charge and both charges reach the surface). Fortunately, these solutions are mostly solved for in Eqs.~\ref{eq:Psurf_0}--\ref{eq:Psurf_3}, except now $P_{\mathrm{surf}}(z_0)$ is replaced with $P_{\mathrm{bulk}}(z_0)$. Making this substitution and solving for the probability distributions gives:

\begin{align}\label{eq:Pbulk_single}
\begin{split}
    P_{0}(E) &= \begin{cases}
    e^{-T_{\mathrm{s}} \cdot E} & 0 \leq E < 1 \\
    0 & \mathrm{else},
    \end{cases} \\
    P_{1}(E) &= \begin{cases}
    \frac{1}{\tau_{\mathrm{CTs}}}e^{-T_{\mathrm{s}} \cdot E}(1-E) & 0 \leq E < 1 \\
    0 & \mathrm{else},
    \end{cases} \\
    P_{2}(E) &= \begin{cases}
    \frac{1}{2\tau_{\mathrm{IIss}}}e^{-T_{\mathrm{s}} \cdot E} E & 0 \leq E < 1 \\[10pt]
    \frac{1}{2\tau_{\mathrm{IIss}}}e^{-T_{\mathrm{s}} \cdot E} (2-E) & 1 \leq E < 2\\
    0 & \mathrm{else},
    \end{cases} \\
    P_{3}(E) &= \begin{cases}
    \frac{e^{T_{\mathrm{o}}(1-E) - T_{\mathrm{s}}} - e^{T_{\mathrm{s}}(1-E) -T_{\mathrm{o}}}}{\tau_{\mathrm{IIso}}(T_{\mathrm{o}} - T_{\mathrm{s}})} & 1 \leq E < 2 ,\, T_{\mathrm{s}} \neq T_{\mathrm{o}} \\[10pt]
    \frac{1}{\tau_{\mathrm{IIso}}}e^{-T_{\mathrm{s}} \cdot E}(2-E) & 1 \leq E < 2 ,\, T_{\mathrm{s}} = T_{\mathrm{o}} \\
    0 & \mathrm{else}.
    \end{cases}
\end{split}
\end{align}

It is evident that the solutions to these problems become complex. One way to determine if these solutions make sense is to examine the boundary conditions. For example, the probability distribution $P_{3}(E)$ in Eq.~\ref{eq:Pbulk_single} ranges from one to two \eh{} pairs of energy. If a charge has an initial position of $z=0$ and immediately creates an opposite charge, the event will produce one \eh{} pair of energy. The same is true if the charge has an initial position of $z=1$ and immediately creates an opposite charge. There is no scenario where this process can produce an energy less than one \eh{} pair. Furthermore, if the charge has an initial position of $z=0$ and creates an opposite charge only when it reaches $z=1$, the event will produce two \eh{} pairs of energy. There is likewise no scenario where this process can produce an energy greater than two \eh{} pairs. The 28 unique solutions found for bulk-single-charge events are catalogued in the Supplemental Material, and the individual solutions are shown together in the middle plot of Fig.~\ref{fig:1eh_solutions}.

\subsection{\label{sssec:bulk_neh_Events}Bulk-\eh{}-Pair Events}
A bulk-\eh{}-pair event starts with the creation of both an electron and hole at some starting position $z_{0}$ that ranges between $z=0$ and $z=1$. As with bulk-single-charge events, we assume that $z_{0}$ is a uniform distribution between the surfaces of the detector and follows Eq.~\ref{eq:Pdist_bulk}. However unlike the solutions for single charges, the solutions for \eh{}-pair events need to keep the distinction between the parameters for electrons and holes. The initial \e{} and \h{} will propagate in opposite directions and are treated as independent charges. The only constraint is the initial starting position that they both share. Like with the single-charge events, the solutions will be the same regardless of which direction of propagation is chosen for the charges. 

Here we will demonstrate how to find the solutions for the simplest scenarios. Let $P_{0}(E)$ be the probability that both the electron and hole reach the surface. We assume the electrons and holes travel toward the $z=1$ and $z=0$ surfaces, respectively. As before, the solution can be found by considering the segments of charge propagation in the scenario. If the total energy of the event is $E$ and the electron contributes an energy of $E_{1}$, then the hole must contribute an energy of $E - E_{1}$. The probability of measuring an energy of $E$ giving a starting position of $z_{0}$ is therefore the probability that the electron contributed an energy of $E_{1}$ starting at $z_{0}$ times the probability that the hole contributed an energy of $E - E_{1}$ starting at $z_{0}$ summed over all possibilities of $E_{1}$. Using the base equations from Eq.~\ref{eq:base_solution_1}, this is written as:

\begin{equation}\label{eq:Peh_0_1}
\begin{split}
    P(E,z_{0}) &= \int_{-\infty}^{\infty}  P^{1}_{\mathrm{S, e}}(E_{1},z_{0}) P^{0}_{\mathrm{S, h}}(E-E_{1},z_{0}) \mathrm{d}E_{1} \\
    &= \int_{-\infty}^{\infty} \delta(E_{1} - 1 + z_{0}) e^{-(1-z_{0})T_{\mathrm{e}}} \times \\
    & \qquad \qquad \delta(E - E_{1} - z_{0}) e^{-z_{0}T_{\mathrm{h}}} \mathrm{d}E_{1} \\
    &= \delta(E-1) e^{-T_{\mathrm{e}} + z_{0}(T_{\mathrm{e}} - T_{\mathrm{h}})}.
\end{split}
\end{equation}

The final step to solve for $P_{0}(E)$ is to multiply Eq.~\ref{eq:Peh_0_1} by Eq.~\ref{eq:Pdist_bulk} and integrate over all $z_{0}$. However this last step must be considered separately for when $T_{\mathrm{e}} = T_{\mathrm{h}}$ and $T_{\mathrm{e}} \neq T_{\mathrm{h}}$ in order to avoid undefined solutions. For the case where $T_{\mathrm{e}} \neq T_{\mathrm{h}}$, $P_{0}(E)$ is found to be:

\begin{equation}\label{eq:Peh_0_2}
\begin{split}
    P_{0}(E) &= \int_{-\infty}^{\infty}  P(E,z_{0}) P_{\mathrm{bulk}}(z_{0}) \mathrm{d}z_{0} \\
    &= \int_{0}^{1} \delta(E-1) e^{-T_{\mathrm{e}} + z_{0}(T_{\mathrm{e}} - T_{\mathrm{h}})} \mathrm{d}z_{0}\\
    &= \frac{1}{T_{\mathrm{e}} - T_{\mathrm{h}}}\delta(E-1) e^{-T_{\mathrm{e}}} \left[ e^{z_{0}(T_{\mathrm{e}} - T_{\mathrm{h}})}\right]_{0}^{1} \\
    &= \delta(E-1) \frac{e^{-T_{\mathrm{h}}} - e^{-T_{\mathrm{e}}}}{T_{\mathrm{e}} - T_{\mathrm{h}}}.
\end{split}
\end{equation}

For the case where $T_{\mathrm{e}} = T_{\mathrm{h}} \equiv T$, $P_{0}(E)$ is found to be:

\begin{equation}\label{eq:Peh_0_3}
\begin{split}
    P_{0}(E) &= \int_{-\infty}^{\infty}  P(E,z_{0}) P_{\mathrm{bulk}}(z_{0}) \mathrm{d}z_{0} \\
    &= \int_{0}^{1} \delta(E-1) e^{-T} \mathrm{d}z_{0}\\
    &= \delta(E-1) e^{-T}.
\end{split}
\end{equation}

The next scenario to consider is when the \e{} is trapped while the \h{} reaches the surface. Let the probability distribution for this process be $P_{1}(E)$. As before, $E_{1}$ is the energy contribution from the electron, and $E - E_{1}$ is the energy contribution from the hole. The probability for measuring an energy $E$ given a starting position of $z_{0}$ is found in the same way as in Eq.~\ref{eq:Peh_0_1}, except that for the electron, the appropriate base equation from Eq.~\ref{eq:base_solution_2} is used:

\begin{align}\label{eq:Peh_1_1}
\begin{split}
    P(E,z_{0}) &= \int_{-\infty}^{\infty}  P^{1}_{\mathrm{CTe}}(E_{1},z_{0}) P^{0}_{\mathrm{S, h}}(E-E_{1},z_{0}) \mathrm{d}E_{1} \\
    &= \int_{0}^{1-z_{0}} \frac{1}{\tau_{\mathrm{CTe}}} e^{-T_{\mathrm{e}} \cdot E_{1}} \times \\
    & \qquad \qquad \delta(E - E_{1} - z_{0}) e^{-z_{0}T_{\mathrm{h}}} \mathrm{d}E_{1} \\
    &= \begin{cases}
    \frac{1}{\tau_{\mathrm{CTe}}} e^{-T_{\mathrm{e}} \cdot E + z_{0}(T_{\mathrm{e}} - T_{\mathrm{h}})} & z_{0} \leq E < 1\\
    0 & \mathrm{else}.
    \end{cases}
\end{split}
\end{align}

Again we can find $P_{1}(E)$ by multiplying Eq.~\ref{eq:Peh_1_1} with Eq.~\ref{eq:Pdist_bulk} and integrating over $z_{0}$. For the case where $T_{\mathrm{e}} \neq T_{\mathrm{h}}$, $P_{1}(E)$ is found to be:

\begin{align}\label{eq:Peh_1_2}
\begin{split}
    P_{1}(E) &= \int_{-\infty}^{\infty}  P(E,z_{0}) P_{\mathrm{bulk}}(z_{0}) \mathrm{d}z_{0} \\
    &= \int_{0}^{1} P(E,z_{0}) \mathrm{d}z_{0}\\
    &= \int_{0}^{E} \frac{1}{\tau_{\mathrm{CTe}}} e^{-T_{\mathrm{e}} \cdot E + z_{0}(T_{\mathrm{e}} - T_{\mathrm{h}})} \mathrm{d}z_{0}\\
    &= \frac{1}{\tau_{\mathrm{CTe}}(T_{\mathrm{e}} - T_{\mathrm{h}})}e^{-T_{\mathrm{e}} \cdot E} \left[ e^{z_{0}(T_{\mathrm{e}} - T_{\mathrm{h}})} \right]_{0}^{E} \\
    &= \begin{cases}
    \frac{e^{-T_{\mathrm{h}} \cdot E} - e^{-T_{\mathrm{e}} \cdot E}}{\tau_{\mathrm{CTe}}(T_{\mathrm{e}} - T_{\mathrm{h}})} & 0 \leq E < 1 \\
    0 & \mathrm{else}.
    \end{cases}
\end{split}
\end{align}

For the case where $T_{\mathrm{e}} = T_{\mathrm{h}} \equiv T$, $P_{1}(E)$ is found to be:

\begin{align}\label{eq:Peh_1_3}
\begin{split}
    P_{1}(E) &= \int_{-\infty}^{\infty}  P(E,z_{0}) P_{\mathrm{bulk}}(z_{0}) \mathrm{d}z_{0} \\
    &= \int_{0}^{1} P(E,z_{0}) \mathrm{d}z_{0}\\
    &= \int_{0}^{E} \frac{1}{\tau_{\mathrm{CTe}}} e^{-T \cdot E} \mathrm{d}z_{0}\\
    &= \frac{1}{\tau_{\mathrm{CTe}}}e^{-T \cdot E} \left[ z_{0} \right]_{0}^{E} \\
    &= \begin{cases}
    \frac{e^{-T \cdot E} \cdot E}{\tau_{\mathrm{CTe}}} & 0 \leq E < 1 \\
    0 & \mathrm{else}.
    \end{cases}
\end{split}
\end{align}

The solutions for the other scenarios can be found by considering the probabilities for the process that happens to each charge and constraining the energy contribution from each charge to the total measured energy. In total there are 16 unique solutions found for bulk-\eh{}-pair events, which are catalogued in the Supplemental Material. The individual solutions are shown together in the bottom plot of Fig.~\ref{fig:1eh_solutions}.

\section{Sub-Peak Distributions due to Non-Ionizing Photons}\label{app:sub_peak}
Section~\ref{ssec:extended_model_surfacetrapping} introduced the phenomenon of non-ionizing energy deposition and how the surface trapping effect can be incorporated into the extended detector response model. In this model, the number of photons that hit the detector is given by a Poisson distribution with a mean of $\lambda$. Each photon will create an \eh{} pair, where there is a probability $\alpha$ that the \eh{} pair undergoes surface trapping. Absorbed photons that result in an \eh{} pair that undergoes surface trapping are classified as non-ionizing photons, whereby the deposited energy in the detector will only be the absorption energy of the photon $E_{\gamma}$. This effect results in the formation of a sub-peak structure at each \eh{}-pair peak in the energy spectrum, as can be seen in the middle plot of Fig.~\ref{fig:surface_trapping_examples}. This appendix provides further details on the distribution of these sub-peak structures and its dependence on $\lambda$ and $\alpha$.

Each sub-peak corresponds to $q$ ionizing photons and $p$ non-ionizing photons. The $q$ ionizing photons will produce $q$ \eh{} pairs that will propagate through the detector where they may undergo bulk CT and II processes. As will be discussed below, the bulk CT and II processes do not affect the shape of the underlying sub-peak distributions, and thus can be ignored. For the sub-peak distribution of $p$ at the $q\textrm{th}$ \eh{}-pair peak, we want to determine the probability $P(p \,| \,q)$, simply given as:
\begin{equation}\label{eq:Pp_given_q}
P(p \,| \,q) = \frac{P(q\,\cap \, p)}{P(q)}.
\end{equation}

To find these probabilities, we must first consider the probabilities of the separate processes. The total number of photons absorbed in the detector is $(p+q)$, and the probability of $(p+q)$ photons hitting the detector is determined from a Poisson distribution with a mean of $\lambda$. The probability of having $p$ non-ionizing photons is determined from a binomial distribution with $(p+q)$ trials and a probability of $\alpha$. Therefore $P(q\,\cap \, p)$ is the probability that $(p+q)$ photons hit the detector and $p$ photons are non-ionizing:
\begin{equation}\label{eq:Pq_cap_p}
\begin{split}
P(q \, \cap \,p) &= \textrm{Poiss.}((p+q); \lambda) \times \textrm{Binom.}(p; (p+q), \alpha) \\
&= \frac{\lambda^{(q+p)}e^{-\lambda}}{(q+p)!} \frac{(q+p)!}{p! \,q!}\alpha^{p}(1-\alpha)^{q} \\
&= \frac{\lambda^{(q+p)}e^{-\lambda}}{p! \,q!} \alpha^{p}(1-\alpha)^{q}.
\end{split}
\end{equation}

If the mean number of photons hitting the detector is $\lambda$ and there is a $(1-\alpha)$ probability that a photon will be ionizing, then the mean number of ionizing photons hitting the detector is $\lambda \cdot (1-\alpha)$. Therefore $P(q)$ is determined from a Poisson distribution with a mean of $\lambda \cdot (1-\alpha)$:
\begin{equation}\label{eq:P_q_photons}
\begin{split}
P(q) &= \textrm{Poiss.}(q; \lambda(1-\alpha)) \\
&= \frac{\lambda^{q}(1-\alpha)^{q}e^{-\lambda(1-\alpha)}}{q!}.
\end{split}
\end{equation}

Likewise, $P(p)$ is determined from a Poisson distribution with a mean of $\lambda \cdot \alpha$:
\begin{equation}\label{eq:P_p_photons}
\begin{split}
P(p) &= \textrm{Poiss.}(p; \lambda \cdot \alpha) \\
&= \frac{\lambda^{p}\alpha^{p}e^{-\lambda \cdot \alpha}}{p!}.
\end{split}
\end{equation}

Using Eqs.~\ref{eq:Pq_cap_p} and~\ref{eq:P_q_photons}, Eq.~\ref{eq:Pp_given_q} becomes:
\begin{equation}\label{eq:Pp_given_q_2}
\begin{split}
P(p \,| \,q) &= \frac{q!}{p! \,q!} \frac{\lambda^{(q+p)}e^{-\lambda} \alpha^{p}(1-\alpha)^{q}}{\lambda^{q}(1-\alpha)^{q}e^{-\lambda(1-\alpha)}} \\
&= \frac{\lambda^{p}\alpha^{p}e^{-\lambda \cdot \alpha}}{p!} \\
&= \textrm{Poiss.}(p; \lambda \cdot \alpha).
\end{split}
\end{equation}

Equation~\ref{eq:Pp_given_q_2} shows that the sub-peak distribution of $p$ for a given $q$ is just the probability of having $p$ non-ionizing photons, which is a Poisson distribution with a mean of $\lambda \cdot \alpha$. Importantly, the sub-peak distribution is \textit{independent} of $q$, and is therefore the same for each \eh{}-pair peak. As shown in Fig.~\ref{fig:surface_trapping_examples}, when the resolution smearing is applied, these sub-peak distributions shift the location of the \eh{}-pair peaks in the spectrum. The amount that the peaks are shifted by $\Delta E_{\textrm{ph}}$ is determined from mean energy of non-ionizing photons at each \eh{}-pair peak. As the sub-peak distributions are the same for each peak, the amount that each peak is shifted by is also constant:
\begin{equation}\label{eq:peak_shift}
\Delta E_{\textrm{ph}} = E_{\gamma} \cdot \lambda \cdot \alpha.
\end{equation}

Lastly, we consider what effect the bulk CT and II processes may have on the sub-peak distributions. For the $q\textrm{th}$ \eh{}-pair peak there are $q$ ionizing photons and thus $q$ \eh{} pairs that propagate through the detector. The peaks in the sub-peak distribution only arise when all of the primary charges from the \eh{} pairs reach the surface without undergoing a CT or II process, otherwise the measured energy will be in a non-quantized region of the spectrum. Appendix~\ref{app:single_eh_solution} showed that the probability of a charge from a surface event to traverse the detector without undergoing a CT or II process is $e^{-T_{\textrm{e/h}}}$, where $T_{\textrm{e/h}}$ encodes the CT and II probabilities for either the electron or hole. The probability of $q$ charges from a surface event to traverse the detector without undergoing CT or II processes is found from a binomial distribution with $q$ trials and a probability of $e^{-T_{\textrm{e/h}}}$: $\textrm{Binom.}(q; q,e^{-T_{\textrm{e/h}}}) = e^{-qT_{\textrm{e/h}}}$. Therefore while the overall scaling of the sub-peak distribution depends on $q$, the shape of the underlying distribution remains constant for each \eh{}-pair peak.

\section{Limitations of the Single-\eh{}-Pair Solutions}\label{app:model_limit}
As discussed in Sec.~\ref{sec:discussion}, the exponential CTII model is limited by the highest order of processes that are modelled. Specifically, solutions for surface events and bulk-single-charge events are found for up to second-order processes, whereas for bulk-\eh{}-pair events, solutions are found for up to first-order processes. In order to assess and quantify these limitations, the single-\eh{}-pair solutions are compared to simple MC simulations of the CT and II processes. The MC simulations model the CT and II processes using the same initial assumptions as the analytical model: that the probability distributions of CT or II occurring are described by Eq.~\ref{eq:Pi_z}, and where CT and II processes are parametrized by the characteristic lengths $\tau_{i}$. However unlike the analytical model, the MC simulations are able to include higher-order CT and II processes. In these MC simulations, there are no physical or detector response processes that are modelled other than CT, II, and generic resolution smearing.

We would like to determine where these higher-order processes become significant such that the analytical model is no longer a suitable description of the MC simulations, and thus of these CT and II processes. There are two main factors that will cause the analytical solutions to deviate from the MC simulations. The first is the total probability of impact ionization, and the second is the total number of events in the simulation. Increasing either the total probability of II or the total number of events will increase the number of events in the MC simulation that undergo higher-order CT or II process that the analytical solutions do not model. 

\begin{figure}[t!]
\begin{center}
\includegraphics[width=1.0\columnwidth]{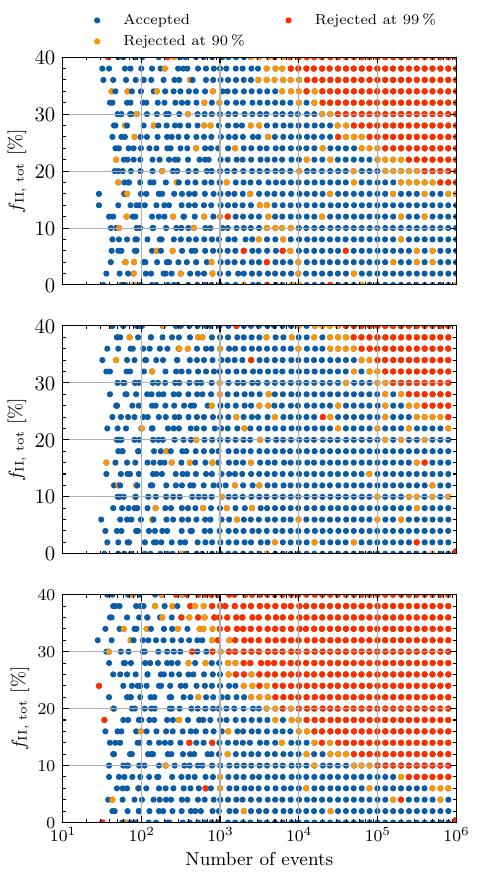}
\end{center}
\caption[KS test results for the single electron-hole-pair solutions]{Results of the KS tests performed using the analytical CT and II solutions and the MC simulations, where each dot corresponds to a test performed for a particular value of the total II probability for a single charge $f_{\textrm{II, tot}}$ and of the total number of events in the simulation. The tests were performed separately for surface events (top), bulk-single-charge events (middle), and bulk-\eh{}-pair events (bottom).}
\label{fig:KS_test_results}
\end{figure}

The limitations of the single-\eh{}-pair solutions can then evaluated by using a simple procedure. For each event type, we scanned over the total II probability in the model and the total number of events in the MC simulation. After computing the analytical model and running the simulation for each set of parameters, we performed a Kolmogorov–Smirnov (KS) test to determine if the simulated spectrum is described by the analytical model for a given level of confidence. For simplicity, we define the total II probability of a single charge $f_{\textrm{II, tot}}$ as $f_{\textrm{II, tot}} = f_{\textrm{IIee}} + f_{\textrm{IIeh}} = f_{\textrm{IIhe}} + f_{\textrm{IIhh}}$, where each $f_{i}$ is equal to $f_{\textrm{II, tot}}/2$. For surface events and bulk-single-charge events, the tests assume that the initial charge is an electron. Furthermore, each of the solutions and simulations assume a small CT probability of $f_{\textrm{CTe}} = f_{\textrm{CTh}} = 1\,\%$ in order to include all possible processes. The KS tests take the null hypothesis that the MC simulation is described by the same probability distribution as the analytical model, and the results from the tests are subsequently placed into three categories: accepted (failed to reject the null hypothesis at 90\,\% confidence level), rejected the null hypothesis at a 90\,\% confidence level, and rejected the null hypothesis at a 99\,\% confidence level. The results of the KS tests for each event type are shown in Fig.~\ref{fig:KS_test_results}.

The test results from Fig.~\ref{fig:KS_test_results} clearly illustrate the regions of this parameter space where the analytical solutions of the exponential CTII model deviate from the MC simulations. For reference, measurements of $f_{\textrm{II, tot}}$ in HVeV detectors have been on the order of 1\,\%~\cite{Ponce2020_2,HVeVR2}. However, these results represent a worst-case scenario for the model, as other parameters can extend the boundary of this limitation. For instance, high CT probabilities will generally lower the probabilities of high-order II processes. Furthermore, these tests were performed using the single-\eh{}-pair solutions, whereas modelling the full detector response will often require the multi-\eh{} solutions. In many cases, the multi-\eh{} solutions will cause the high-order II processes to be subdominant within the total probability distribution, as can be seen by comparing the top and bottom plots of Fig.~\ref{fig:single-multi-CTII}. In these scenarios, the analytical solutions may adequately describe the CT and II processes even for higher II probabilities or for a larger number of events.

\clearpage

\bibliography{refs}

\end{document}

%% file: authors.tex
\author{M.~J.~Wilson}\email{matthew.wilson@kit.edu} \affiliation{Institut f{\"u}r Astroteilchenphysik, Karlsruher Institut f{\"u}r Technologie,  76133  Karlsruhe, Germany}
\author{A.~Zaytsev}\email{alexander.zaytsev@kit.edu} \affiliation{Institut f{\"u}r Astroteilchenphysik, Karlsruher Institut f{\"u}r Technologie,  76133  Karlsruhe, Germany}
\author{B.~von~Krosigk} \affiliation{Kirchhoff-Institut f{\"u}r Physik, Universit{\"a}t Heidelberg, 69117 Heidelberg, Germany}

\author{I.~Alkhatib} \affiliation{Department of Physics, University of Toronto, Toronto, ON M5S 1A7, Canada}
\author{M.~Buchanan} \affiliation{Department of Physics, University of Toronto, Toronto, ON M5S 1A7, Canada}
\author{R.~Chen} \affiliation{Department of Physics \& Astronomy, Northwestern University, Evanston, IL 60208-3112, USA}
\author{M.D.~Diamond} \affiliation{Department of Physics, University of Toronto, Toronto, ON M5S 1A7, Canada}
\author{E.~Figueroa-Feliciano} \affiliation{Department of Physics \& Astronomy, Northwestern University, Evanston, IL 60208-3112, USA}
\author{S.A.S.~Harms} \affiliation{Department of Physics, University of Toronto, Toronto, ON M5S 1A7, Canada}
\author{Z.~Hong} \affiliation{Department of Physics, University of Toronto, Toronto, ON M5S 1A7, Canada}
\author{K.T.~Kennard} \affiliation{Department of Physics \& Astronomy, Northwestern University, Evanston, IL 60208-3112, USA}
\author{N.A.~Kurinsky} \affiliation{SLAC National Accelerator Laboratory/Kavli Institute for Particle Astrophysics and Cosmology, Menlo Park, CA 94025, USA}
\author{R.~Mahapatra} \affiliation{Department of Physics and Astronomy, and the Mitchell Institute for Fundamental Physics and Astronomy, Texas A\&M University, College Station, TX 77843, USA}
\author{N.~Mirabolfathi} \affiliation{Department of Physics and Astronomy, and the Mitchell Institute for Fundamental Physics and Astronomy, Texas A\&M University, College Station, TX 77843, USA}
\author{V.~Novati} \thanks{Presently at LPSC, CNRS, Universit{\'e} Grenoble Alpes, Grenoble, France} \affiliation{Department of Physics \& Astronomy, Northwestern University, Evanston, IL 60208-3112, USA}
\author{M.~Platt} \affiliation{Department of Physics and Astronomy, and the Mitchell Institute for Fundamental Physics and Astronomy, Texas A\&M University, College Station, TX 77843, USA}
\author{R.~Ren} \affiliation{Department of Physics, University of Toronto, Toronto, ON M5S 1A7, Canada}
\author{A.~Sattari} \affiliation{Department of Physics, University of Toronto, Toronto, ON M5S 1A7, Canada}
\author{B.~Schmidt} \thanks{Presently at IRFU, CEA, Universit{\'e} Paris-Saclay, France} \affiliation{Department of Physics \& Astronomy, Northwestern University, Evanston, IL 60208-3112, USA}
\author{Y.~Wang} \affiliation{Department of Physics, University of Toronto, Toronto, ON M5S 1A7, Canada}
\author{S.~Zatschler} \thanks{Presently at LPSC, CNRS, Universit{\'e} Grenoble Alpes, Grenoble, France} \affiliation{Department of Physics, University of Toronto, Toronto, ON M5S 1A7, Canada}
\author{E.~Zhang} \affiliation{Department of Physics, University of Toronto, Toronto, ON M5S 1A7, Canada}
\author{A.~Zuniga} \affiliation{Department of Physics, University of Toronto, Toronto, ON M5S 1A7, Canada}